\documentclass[preprint2]{aastex}

\usepackage{graphicx}
\usepackage{color}
\usepackage{xspace}
\usepackage{amsmath}

\newcommand{\be}{\begin {equation}}
\newcommand{\ee}{\end {equation}}
\newcommand{\ltsima}{$\; \buildrel < \over \sim \;$}
\newcommand{\simlt}{\lower.5ex\hbox{\ltsima}}
\newcommand{\gtsima}{$\; \buildrel > \over \sim \;$}
\newcommand{\simgt}{\lower.5ex\hbox{\gtsima}}

\shortauthors{Valdarnini R.}

\begin{document}


\title{ Improved Performances in Subsonic Flows of an SPH \\
   Scheme with Gradients Estimated using  an Integral Approach }


\author{R. Valdarnini$^{1,2}$ }
\affil{$^1$SISSA, Via Bonomea 265, I-34136, Trieste, Italy}
\affil{$^2$Iniziativa Specifica QGSKY, Via Valerio 2, I-34127 Trieste, 
Italy}

\email{valda@sissa.it}



\begin{abstract}
In this paper we present results from a series of hydrodynamical tests
aimed at validating the performance  of a 
smoothed particle hydrodynamics (SPH) formulation in which gradients are 
derived  from an integral approach.
We specifically investigate the code behavior with subsonic flows, where it is  
well known that  zeroth-order inconsistencies  present in standard SPH 
make it particularly problematic to correctly model the fluid dynamics.

In particular we consider the Gresho-Chan vortex problem, 
the growth of Kelvin-Helmholtz instabilities, 
the statistics of driven subsonic turbulence and  the
cold Keplerian disc problem.
We compare simulation results for the different tests with those obtained,
for the same initial conditions, using standard SPH.
We also compare the results with the corresponding ones obtained previously 
with other numerical methods, such as codes based on a 
moving-mesh scheme or Godunov-type Lagrangian meshless methods.

We quantify code performances by introducing error norms and spectral properties 
of the particle distribution, in a way similar to what was done in other works.
We find that the new SPH formulation exhibits strongly reduced gradient errors and 
outperforms standard SPH in all of the tests considered.
In fact, in terms of accuracy we find  good agreement between the simulation 
results of the new scheme and those produced using other recently proposed 
numerical schemes.
These findings suggest that the proposed method can be successfully applied 
for many astrophysical problems in which the presence of subsonic flows
previously limited the use of SPH, with the new scheme now being 
competitive in these regimes with other numerical methods.

\end{abstract}




\section{Introduction}

Application of computational fluid dynamics to many astrophysical problems
has grown steadily over the years  with advances in computational 
power and it has now become a standard tool for studying the 
non-linear evolution of baryonic structures in the Universe.

There are two methods commonly used in numerical astrophysics  for
solving the Euler equation. The first method makes use of a spatial grid,
 either fixed \citep{st92,no99,sto08} or adaptative 
\citep{fr00,te02,no05,br14}.
The second method is a Lagrangian mesh-free numerical scheme, known as 
smoothed particle hydrodynamics \citep[SPH:][]{lu77,gm77}
  in which particles are used to model fluid properties.
Both methods have advantages and weaknesses which are 
specific to the numerical approach on which each method is based.

Because of its Lagrangian nature, SPH possesses very good conservation
properties, moreover the method is free of advection errors present 
in mesh codes and is naturally adaptative because particle trajectories trace 
the mass. This latter feature is particularly useful in many astrophysical 
problems involving collapse of the structure  under study.

The method, however, is not free from significant drawbacks and 
more specifically it has been found that in several hydrodynamical test cases
there are significant differences between the results obtained using 
the two methods, with SPH failing to properly model the correct behavior
\citep{os05,ag07,wa08,ta08,mi09,read10,vrd10,ju10}.

More specifically, \citet{ag07} found that SPH fails to resolve the 
formation of Kelvin-Helmholtz (KH) instabilities at fluid interfaces.
This is strongly related to the fluid mixing properties of SPH as well as
 to the 
lack of a core entropy in non-radiative simulations of galaxy clusters,
 in contrast  with what is found using mesh-based codes \citep{wa08,mi09}.

It is now widely recognized that  the origin of these errors is due 
to two distinct problems which are present in SPH.
{
 The first problem originates from the inconsistencies of standard SPH 
when dealing with 
 steep density gradients at  
contact discontinuities, the so-called local mixing instability 
\citep{read10},
thereby suppressing the growth 
of KH instabilities at the fluid interfaces.
}
The second problem is inherent in the discrete nature of SPH, in which 
 a finite set of particles is used to model the fluid.
The discretization implies the presence of a zeroth-order error in the 
momentum equation due to sampling effects \citep{read10}, the so-called 
$E_0$ error. This error can be 
reduced if one increases the number of neighbors $N_n$ present within 
the kernel, but for the standard cubic spline kernel there is 
a threshold value for $N_n$ beyond which a clumping instability develops 
thus degrading the convergence rate.

In view of the benefits of the SPH method previously outlined, there have been 
many attempts to eliminate or reduce these difficulties.
Several solutions have been proposed, concerning 
the problem posed by the standard formulation at contact 
discontinuities.

One solution is to modify the equations, so that it is the pressure rather 
than the density which is smoothed \citep{ri01,read10,ho13,sa13,hu14}.
 On the other hand, \citet{pr08} proposed to include in the SPH energy equation
a term of artificial conductivity (AC)  
with the aim of smoothing the thermal energy across fluid interfaces 
and thus removing the associated entropy gap.  
This approach is similar to that used by \citet{wa08}, who mimicked the 
effect
 of subgrid turbulence by adding a heat diffusion term to the 
equations. 
The method, however, requires some care in the implementation of the 
conduction switches to avoid the risk of getting too much diffusion.

By performing a suite of hydrodynamical tests 
\citep[][ hereafter V12]{va12} it has 
been found that the method yields consistent results when 
contrasted with those obtained using mesh-based codes.
In particular \citep[V12;][]{bi15},  the level of core 
entropies produced in  simulations of non-radiative galaxy clusters are now
comparable with those of grid codes. Recently \citet{be16}, have
proposed a modification to the standard SPH code Gadget-II \citep{sp05}
which incorporate the new AC term into the hydrodynamic equations.

Finally, other variants of  SPH are based on Riemann solvers 
 \citep[Godunov-SPH:][]{I02,ch10,mu11},
 Voronoi tessellation techniques \citep{hs10} or on the  use of 
high order dissipation switches \citep{read12}.

The zeroth-order inconsistency is due to the inability of the SPH method
to properly reproduce a constant function because of finite resolution
 \citep{di99,li03}, thus leading to poor gradient estimates 
\citep{read10,na12} 
and in turn affecting the momentum equations.
Keeping these errors under control becomes problematic when 
dealing with subsonic flows, as in the case of subsonic turbulence 
\citep{ba12} or with Rayleigh-Taylor instabilities \citep[][V12]{ab11,ga12}. 

A possible solution is to drastically increase the number of neighbors
used in the simulation. In this case, the clumping instability  can be avoided
 either by  modifying the shape of the cubic spline kernel 
\citep{read10}, or by adopting \citep{de12} the 
Wendland kernels \citep{we95}. These kernels are characterized by the 
specific property of not being subject  to clumping instabilities
in the large $N_n$ limit.

Another possibility is to consider other  discretizations of the
 momentum equation \citep{mo96,ab11} but this comes at the cost of 
losing energy and momentum conservation \citep{pr12}, thus making the 
scheme of little use in practice.

Finally, to overcome the difficulties of SPH mentioned above, new 
numerical schemes have been proposed \citep{sp10,du11,ho15,sa15,pa16}
which  aim to retain the advantages of using both SPH and mesh-based codes.
These new schemes are quite numerical complex and,
in some cases,  their space discretization does not seem to be optimal as 
required by forthcoming parallel computing systems consisting of several 
million cores \citep{sa15}.

A satisfactory solution to the problem  of zeroth-order inconsistency 
in SPH has been presented by \citet{ga12}, who showed how the accuracy in gradient
estimates can be greatly improved by calculating first order derivatives 
by means of  the evaluation of integrals and the use of matrix inversions.
The resulting tensor scheme has been tested in a variety of hydrodynamical 
test cases \citep{ga12,ro15}, showing significant improvements as compared
with the standard formulation. A crucial feature of the method is that 
 it retains the Lagrangian nature of SPH, unlike previous attempts aimed 
at improving gradient accuracy.

Motivated by these findings we here further investigate the performance of 
the new scheme, paying particular attention to its behavior in the regime
of subsonic flows, where it has been found that standard SPH presents its
major difficulties.

The goal of this paper is to demonstrate that, for the hydrodynamical tests 
considered here, the new SPH formulation gives results with 
accuracy comparable to that of mesh-based codes.
Thus, the new code can be profitably used for many astrophysical problems 
without the shortcomings of standard SPH. The main advantage of the new scheme 
is that  it keeps its fully Lagrangian nature, while retaining a relative 
simplicity 
in its implementation as compared with new numerical schemes recently proposed.

The  paper is organized as follows. In Section \ref{sec:hydro} we present the 
hydrodynamical method and the implementation of the integral-based approach. 
Some basic properties of the most widely used SPH kernels are briefly
 reviewed in Section \ref{sec:ker}.
 The results of the hydrodynamical tests are given in Section \ref{sec:tests},
where we consider the Gresho-Chan vortex problem, 
the development of KH instabilities, 
the statistic of driven subsonic turbulence and  finally the
Keplerian disc problem.
Our main results and conclusions are summarized in Section \ref{sec:conc}.

\section{Hydrodynamic  method } \label{sec:hydro}
This section reviews the basic features of SPH; for a comprehensive
review  see  \citet{ro09} and  \citet{pr12}.

\subsection{Basic equations} \label{subsec:method}

In SPH, the fluid is described within the solution domain by a set of
$N$ particles with mass $m_i$, velocity $\vec v_i$, density $\rho_i$,
and specific entropy $A_i$ ( we use the convention of having 
Latin indices denoting particles and Greek indices denoting 
the spatial dimensions). Here, we integrate the entropy per particle
\citep{SH02} in place of the thermal energy per unit mass $u_i$ 
\citep{hk89,wa04}.
 The entropy $A_i$ is
related to the particle pressure $P_i$ by
$P_i=A_i\rho_i^{\gamma}=(\gamma-1) \rho_i u_i$, where $\gamma=5/3$ for
a mono-atomic gas.  The density estimate at the particle position 
$\vec r_i$ is given by
 \begin{equation}
 \rho_i=\sum_j m_j W(|\vec r_{ij}|,h_i),
    \label{rho.eq}
 \end{equation}
where $W(|\vec r_i-\vec r_j|,h_i)$ is the interpolating kernel
 which is zero for $|\vec r_i-\vec r_j|\geq\zeta h_i$ \citep{pr12}.
Since the kernel has compact support, the sum in Equation (\ref{rho.eq}) 
is over a finite number of particles.
The smoothing length $h_i$  is implicitly defined  by
 \begin{equation}
h_i=\eta (m_i/\rho_i)^{1/D}~,
  \label{hzeta.eq}
 \end{equation}
so that in two and three dimensions, respectively, 
$N^{2D}_{nn}= {\pi (\zeta \eta)^2 }$ and
$N^{3D}_{nn}= {4 \pi (\zeta \eta)^3 }/{3}$ 
are the {\bf mean} number of neighboring particles  
of particle $i$ within a radius
$\zeta h_i$.
{ 
In principle, for a given parameter $\eta$, the solution of Equation \ref{hzeta.eq}
allows for non-integer values of $N_{n}$ \citep{pr12}. Here we solve
the equation for the $h_i$ by requiring an integer value 
for $N_{nn}$, to which we will generically refer in the following as the 
neighbor number.
 }

Following \citet{pr12} the Euler equations are  derived using a 
Lagrangian formulation
   \begin{equation}
  \frac {d \vec v_i}{dt}=-\sum_j m_j \left[
  \frac{P_i}{\Omega_i \rho_i^2}
  \vec \nabla_i W_{ij}(h_i) +\frac{P_j}{\Omega_j \rho_j^2}
   \vec \nabla_i W_{ij}(h_j)
\right]~,
  \label{fsph.eq}
   \end{equation}
where  $\Omega_i$ is defined as 
   \begin{equation}
   \Omega_i=\left[1-\frac{\partial h_i}{\partial \rho_i}
   \sum_k m_k \frac{\partial W_{ik}(h_i)}{\partial h_i}\right]~.
    \label{fh.eq}
   \end{equation}

\subsection{Artificial viscosity } \label{subsec:visco}
The  momentum equation (\ref{fsph.eq}) must be generalized to include an 
artificial viscosity (AV) term which in SPH represents the effects of shocks.
 This is introduced in order  to prevent particle streaming  and 
convert kinetic energy into heat  at shocks; the new term reads

   \begin{equation}
   \left (\frac {d \vec v_i}{dt}\right )_{AV}=-\sum_i m_j \Pi_{ij} \vec \nabla_i \bar W_{ij}~,
    \label{fvis.eq}
   \end{equation}
   where $\bar W_{ij}=
   \frac{1}{2}(W(r_{ij},h_i)+W(r_{ij},h_j))$ is the symmetrized kernel
   and $\Pi_{ij}$ is the AV tensor.

 The latter is written following the formulation of \citet{mo97},
based on an analogy with the Riemann problem :
   \begin{equation}
\Pi_{ij} =
 -\frac{\alpha_{ij}}{2} \frac{v^{AV}_{ij} \mu_{ij}} {\rho_{ij}} f_{ij}~,
  \label{pvis.eq}
 \end{equation}
where $\rho_{ij}$ is the average density, 
  $\mu_{ij}= \vec v_{ij} \cdot
 \vec r_{ij}/|r_{ij}|$ if $ \vec v_{ij} \cdot \vec r_{ij}<0$ but zero
 otherwise and $\vec v_{ij}= \vec v_i - \vec v_j$. The signal velocity 
$v^{AV}_{ij}$ is introduced  as
   \begin{equation}
v^{AV}_{ij}= c_i +c_j - 3 \mu_{ij}~,
  \label{vsig.eq}
 \end{equation}
 with $c_i$ being the sound velocity. The amount of AV is regulated
by the parameter $\alpha_i$, and $f_i$ is a viscosity limiter 
 which is introduced so as to suppress  AV when strong 
shear flows are present. 
This is written as  \citep{ba95} 
   \begin{equation}
  f_i=\frac {|\vec \nabla \cdot \vec v|_i}
  {|\vec \nabla \cdot \vec v|_i+|\vec \nabla \times \vec v|_i}~,
   \label{fdamp.eq}
 \end{equation}
 where $(\vec \nabla \cdot \vec v)_i$ and $(\vec \nabla \times \vec
 v)_i$ are estimated according to the SPH formalism.

The early SPH formulation \citep{mo05},  assumed a constant
viscosity parameter $\alpha_i$ of order unity for all the particles, 
thus making the scheme excessively viscous away from shocks.
In the literature, the SPH scheme with this viscosity parametrization is 
often referred to as standard SPH,  whereas here we use  this term to indicate 
the SPH formulation which uses the AV switch which now we will describe.

To reduce the amount of AV away from shocks
 \cite{mm97} proposed letting the $\alpha_i$'s
  vary with time according to some source term $S_i$.
The time-evolution of $\alpha_i$ is given by
\begin{equation}
 \frac {d \alpha_i}{dt} =-\frac{\alpha_i-\alpha_{min}}{\tau_i} +{S}_i~,
  \label{alfa.eq}
\end{equation}
where $\alpha_{min}$ is a floor value and 
\begin{equation}
  \tau_i=\frac{h_i}{c_i ~l_d}
    \label{tau.eq}
  \end{equation}
  is a decay time scale which is controlled by the dimensionless decay
  parameter $l_d$. 
  The source term $S_i$ is constructed so that  it increases whenever
$\vec \nabla \cdot \vec v_i <0$ \citep{mm97};  here we adopt a slightly 
modified form 
\citep{va11} which reads 
  \begin{eqnarray*}
 \lefteqn{{\tilde S}_i=} & 
   ~~~f_i S_0 {max}\big(-\big(\vec \nabla \cdot \vec v\big)_i,0\big)
(\alpha_{max}-\alpha_i)  \\
  & \equiv S_i (\alpha_{max}-\alpha_i).
    \label{salfa.eq}
   \end{eqnarray*}

  where $\alpha_{max}$ sets an upper limit  and $S_0$ is unity for $\gamma=5/3$.
 In the following, unless otherwise specified, 
 we adopt a time-dependent AV scheme with parameters 
   $\{\alpha_{min},\alpha_{max},l_d\} = \{0.1,1.5,0.2\}$.
 This set of parameters will be denoted as AV$_2$ 
\cite[See Table 1 of ][ to which we refer for more details]{va11}.

To suppress AV more efficiently  away from shocks, the time dependent AV 
scheme has been further improved by 
\citet{cul10}.
They introduced  the time derivative of 
$\vec \nabla \cdot \vec v_i $  to detect in advance when a flow is 
converging,  as well as higher order gradient estimators and a more sophisticated 
functional form for the viscosity limiter in shear flows.
{ 
The \citet{cul10} scheme will be used in some  test cases
and we refer to the
authors' paper for a detailed description of its implementation in SPH.
}

%
\subsection{The artificial conductivity scheme}\label{subsec:ac}
In the  entropy formulation of SPH, the rate of  
 entropy generation is given by \citep{SH02}
   \begin{equation}
  \frac {d A_i}{dt} =\frac{\gamma-1}{\rho_i^{\gamma-1}}\{
   Q_{AV} +Q_{AC}\}~,
    \label{aen.eq}
   \end{equation}
where the terms in brackets denote different sources ( in the hydrodynamic 
test cases presented here, radiative losses are not considered).
 
 The term  $Q_{AV}$ refers to the numerical viscosity:
   \begin{equation}
  Q_{AV} =  \left ( \frac {d u_i}{dt} \right)_{AV} =
  \frac{1}{2}
  \sum_j m_j \Pi_{ij} \vec v_{ij}\cdot \nabla_i \bar W_{ij}~.
    \label{avis.eq}
   \end{equation}

 The AC term $Q_{AC}$ for the dissipation of energy takes the form
   \begin{equation}
  \left ( \frac {d u_i}{dt} \right)_{AC} =
\sum_j \frac{m_j v^{AC}_{ij}}{\rho_{ij}}
\left[ \alpha^C_{ij}(u_i-u_j) \right ] \vec {e_{ij}}\cdot \vec {\nabla_i}
 \bar W_{ij}~,
  \label{duc.eq}
   \end{equation}
where $ v^{AC}_{ij}$ is the AC signal velocity,
$\vec e_{ij} \equiv \vec r_{ij}/r_{ij}$, and $\alpha^C_{i}$ is the AC parameter 
which is of order unity.
{ 
 The above Equation represents the SPH analogue of a diffusion
equation of the form \citep{pr08}

 \be
  \left ( \frac {d u_i}{dt} \right)_{AC}  \simeq D^{AC}_i \nabla^2 u_i~,
  \label{dudis.eq}
 \ee

where  $D^{AC}_i$ is a numerical heat-diffusion coefficient
given by 

\be
 D^{AC}_i \simeq \frac{1}{2}\alpha^C_{i} v^{AC}_{ij} r_{ij}~.
 \label{dac.eq}
\ee
}

A crucial issue concerns reducing the AC in the absence of contact 
discontinuities. In analogy with the AV scheme, one can define an AC switch with a source term 
given by 
 \begin{equation}
{S^C}_i= f_C h_i\frac{|\nabla^2 u_i|}{\sqrt{u_i+\varepsilon}}
\left(\alpha^C_{max}-\alpha^C_i\right),
    \label{salfac.eq}
 \end{equation}

where the Laplacian of the thermal energy is calculated as done by \cite{br85}
 \begin{equation}
\nabla^2 u_i=2\sum_j m_j \frac{u_i-u_j}{\rho_j}\frac{\vec e_{ij}\cdot
\vec {\nabla} W_{ij}}{r_{ij}}~,
 \label{udii.eq}
   \end{equation}

and for  the signal velocity  we use (V12)
 \begin{equation}
v^{AC}_{ij} = |(\vec v_i-\vec v_j)\cdot \vec r_{ij}|/r_{ij}~.
 \label{vsgv.eq}
 \end{equation}

For the other parameters we set $f_C=1$, $\alpha^C_{min}=0$,   
$\alpha^C_{max}=1.5$, and $\varepsilon=10^{-4} u_i$. 

The time evolution of the AC parameter  $\alpha^C_i$ is similar to that of the AV 

 \begin{equation}
\frac {d \alpha^C_i}{dt} =-\frac{\alpha^C_i-\alpha^C_{min}}{\tau^C_i} +{S^C}_i~,
    \label{alfac.eq}
 \end{equation}
where $\tau^C_i={h_i}/{0.2 c_i }$ sets the decaying time scale away 
from jumps in thermal energy.

{ 
The AC term has been introduced with the purpose of
smoothing the thermal energy at contact discontinuities \citep{pr08}, 
and when using the signal velocity (\ref{vsgv.eq})
can be interpreted as a subgrid model  mimicking  the 
effects of diffusion due to turbulence \citep{wa08}.
Finally, it must be stressed that for the hydrodynamic test problems 
considered here, with the exception of the KH tests, 
thermal energy gradients are null or very small. For these tests 
the impact of AC on simulation results can then be considered negligible.
}

%

\subsection{The Integral Approximation scheme} \label{subsec:IAD}
In SPH, the errors associated with finite sampling cannot be simply 
eliminated by a more accurate interpolation scheme. 
A gradient estimator which is exact at 
linear order can be constructed by using matrix inversion \citep{pr12}, but this 
comes at the cost of losing the conservation properties of SPH.

To avoid these difficulties \citet{ga12} proposed a novel approach in which
SPH first-order derivatives are estimated through the use of integrals. 
This makes the method much less noisy  than in the standard formulation, with 
 accuracy in estimated gradients being greatly improved \citep{ga12,ro15}.

Moreover, a significant benefit of the method is that it retains the Lagrangian 
nature of SPH, thereby ensuring exact conservation of linear and angular 
momentum. 

After the paper of \citet{ga12}, the performance of the scheme  was
investigated in detail by \citet{ro15}; here we briefly describe the 
essential features of the method.
  
Let us define the integral 

   \begin{equation}
 {I(\vec r) }= \int_V \left [ f(\vec r ^{\prime}) -f (\vec r ) \right ]
(\vec r ^{\prime} - \vec r ) W( |\vec r ^{\prime} - \vec r |, h ) d^3 r^{\prime}~,
    \label{iad.eq}
   \end{equation}

where $W$ is  a spherically symmetric and normalized kernel.
By Taylor expanding $f(\vec r^{\prime})$ to first order around $\vec r $ 
 
   \begin{equation}
 {I}_{\alpha} \simeq \vec \nabla_{\beta}  f 
\int \Delta_{\alpha} \Delta_{\beta } W  d^3 r^{\prime}~,
   \label{iadb.eq}
   \end{equation}

where we have introduced the notation 
$\Delta_{\alpha}=(\vec r ^{\prime} - \vec r)_{\alpha}$.
The gradient of the function $f$ is then given by

   \begin{equation}
 \vec \nabla_{\alpha}  f =
  \left [ \tau \right ] ^{-1} _{\alpha\beta} I_{\beta}~,  
    \label{iadc.eq}
   \end{equation}

where $\mathcal{T}=  \{ \tau\} _{\alpha \beta}  $ and

 \begin{equation}
 \tau _{\alpha\beta}=\tau _{\beta \alpha}= \int \Delta_{\alpha} \Delta_{\beta } W  d^3 r^{\prime}~.  
\label{iadd.eq}
 \end{equation}

In SPH integrals are replaced by summations over particles, so that for 
the matrix  $\mathcal{T}$ of particle $i$ one has

   \begin{equation}
   \tau _{\alpha\beta}(i)=\sum_k \frac{m_k}{\rho_k} 
\Delta_{\alpha} ^{ki} \Delta_{\beta }^{ki}  W(r_{ik},h_i)~.    
    \label{iade.eq}
   \end{equation}

In evaluating the discrete equivalent of the integral (\ref{iad.eq}), 
 a key step is to assume that the condition 

\begin{equation}
 \sum_k \frac{m_k}{\rho_k} (\vec r_k-\vec r_i )  W_{ik}  \simeq 0
 \label{sumc.eq}
\end{equation}

is fulfilled  with a certain degree of accuracy. 
In such a case the integral  (\ref{iad.eq}) then becomes 

   \begin{equation}
{I}_{\beta}(i) = \sum_k \frac{m_k}{\rho_k} f_k \Delta_{\beta }^{ki}
  W(r_{ik},h_i)~.   
    \label{iadb2.eq}
   \end{equation}

Because of the approximation (\ref{sumc.eq}), for linear functions 
gradient estimates are no longer exact. 
However it is can easily be seen \citep{ga12} that the gradient approximation 
(\ref{iadc.eq}), obtained using Equations (\ref{iade.eq})  and (\ref{iadb2.eq}), 
is now antisymmetric in the pair $ij$. 
Thus, the condition (\ref{sumc.eq}) in the new scheme is crucial for ensuring 
exact conservation properties.

How well the approximation (\ref{sumc.eq}) is valid depends on the 
particle distribution within the kernel radius. The validity of the  new
scheme has been carefully tested \citep{ga12,ro15} for several 
hydrodynamical problems, demonstrating significant improvements in 
the accuracy of the results with respect to standard SPH.

To summarize , the integral approximation implies the replacement of 
$\left[ \vec \nabla_i W_{ik} \right]_{\alpha} $ in the SPH equations
according to the following prescriptions:

 \begin{equation}
 \left [ \nabla_i  W_{ik}(h_i) \right] _{\alpha}  \rightarrow
\sum_{\beta} C_{\alpha\beta } (i) \Delta_{\beta }^{ki}  W(r_{ik},h_i)~,  
  \label{iadf.eq}
 \end{equation}
 
and

 \begin{equation}
 \left [ \nabla_i  W_{ik}(h_k) \right] _{\alpha}  \rightarrow
\sum_{\beta} C_{\alpha\beta } (k) \Delta_{\beta }^{ki}  W(r_{ik},h_k)~. 
 \label{iadg.eq}
 \end{equation}
 
where $\mathcal{C}=\mathcal{T}^{-1}$ .
In the following, the SPH formulation in which gradients are estimated 
according to the numerical scheme described here, 
  will be referred to as the integral approximation (IA).




\begin{figure}[t]
\includegraphics[height=8cm,width=8cm]{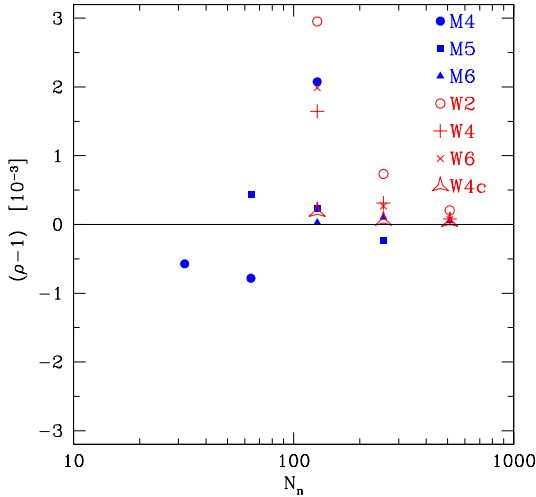}
\caption{ Average SPH density  calculated for a glass-like 
configuration of $128^3$ particles in a periodic box of unit
length. The theoretical expected value is $\rho=1$ and the quantity 
plotted is $(\rho-1)/10^{-3}$. 
Different symbols refer to different kernels;
for the sake of clarity error bars are not shown.
The symbol $W4c$ refers  the Wendland kernel $W4$, but with the 
self-correction term of  \citet[][ Eq. 18]{de12} included.
\label{fig:rho}}
\end{figure}

\section{KERNELS} \label{sec:ker}

In this section  we briefly review some properties of the most
commonly used kernels in SPH. All of these kernels are characterized 
by the property of having compact support and of being continuous up to 
some degree.

A class of kernels which has often been considered in SPH  is that of
the  B-spline functions \citep{pr12}, which are generated via the 1D Fourier
transform:

\be 
M_n(x,h)=\frac{1}{2\pi} \int_{-\infty}^{\infty} \left [ \frac {\sin (kh/2)}
{kh/2} \right] ^n \cos(kx) dk~.
\label{mnke.eq}
\end{equation}

The degree of smoothness increases with $n$ and the kernel approaches the 
Gaussian in the limit $n\rightarrow \infty$. The function $M_n$ is a
polynomial of degree $n-1$ and its derivative is continuously differentiable 
$n-2$ times. By requiring in SPH the continuity of the first and second 
derivative, the first useful kernel is then $M_4$ (cubic spline):

\be
w(q) = \frac{\sigma}{h^D} \left\{ \begin{array}{ll}
\frac{1}{4}{(2-q)}^3-{(1-q)}^3 & 0 \le q < 1,\\
\frac{1}{4}{(2-q)}^3  & 1 \le q < 2, \\
0. &  q \ge 2, 
\end{array} \right. 
\label{quartke.eq}
\ee

where the kernel is non zero for $0\leq q \leq \zeta=2$ and
$\zeta$ is the truncation radius in units of $h_i$. The 
normalization constant takes the values 
 $\sigma={10}/{7\pi} ,{1}/{\pi}$ for $D=2$ and $D=3$,  respectively.

The B-splines next in order which have been considered are $M_5$ 
($\zeta=2.5$) and $M_6$ ( $\zeta=3$); for a more detailed description
of these kernels we refer the reader to \citet{pr12}. 

The stability properties of the $M_n$ kernel family  have been 
investigated by a number of authors \citep[][V12]{mo96,borv04,read10,de12}.
A crucial result which emerges from these analyses is that all of the 
B-splines suffer from pairing instability. 
The number of neighboring particles for which the instability develops
depends on the kernel degree and in 3D lies in the range between 
$N_n\simeq 50$ for $M_4$ up to $N_n\simeq 200$ when the $M_6$ kernel
is used \citep{de12}.

It has been suggested that particle clumping can be avoided by modifying
the kernel shape in order to have a non-zero gradient at the origin.
Examples of this family of kernels are the core-triangle 
\citep[CRT;][]{read10} and the linear quartic \citep[LIQ;][]{vrd10}.
However, such adjustments reflect negatively on the density estimation
ability of the kernels. By introducing a non-zero central derivative, 
the kernel profile becomes steeper and this in turn implies an overestimate
of density when compared with the corresponding B-spline \citep[V12,][]{ro15}.

These attempts to fix the pairing instability by introducing ad hoc 
modifications in the kernel shape have recently been superseeded by 
a new class of kernels. It has been shown \citep{de12} that a necessary
condition for avoiding pairing instability is that of having kernels with 
non-negative Fourier transforms.
A class of kernels with satisfies this property  and has compact support 
are the \citet{we95} functions.
An example in 3D of these functions is the Wendland $C^4$ :

\be
w(q) = \frac{495}{32 \pi} {(1-q)}^6 ( 1+6q + \frac{35}{3} q^2)~,
\label{wedke.eq}
\ee

where $w(q)=0$ if $q>1$. Hereafter we will refer to this kernel as $W4$.
Other classes of Wendland kernels are $C^2$ ($W2$) and  $C^6$ ($W6$).
 We refer to Table 1 of \citet{de12} for the functional forms and normalization
of these kernels. Finally, \citet{ga14} proposed using the sinc functions 
as another class of kernels which can be used to avoid pairing instability.

The accuracy of density estimation in SPH for different kernel families 
has been assessed by many authors 
(Dehnen \& Aly 2012; V12; Rosswog 2015; Zhu et al. 2015).
Here we measure the mean SPH density of $N=128^3$ particles using a glass-like 
particle distribution inside a cube of sidelength unity and total mass one.
Figure \ref{fig:rho} shows the mean SPH density of the particles 
as a function of the neighbor number $N_n$
for different B-splines and Wendland kernels.
 The value of $N_n$ ranges in powers of two between $N_n=32$ and $N_n=512$, 
 with three distinct values of $N_n$ being considered for each  kernel
according to its order ( see Figure \ref{fig:rho} ).
To avoid overcrowding in the plot, standard deviation $\sigma$'s are not shown, 
but the general tendency is of  $\sigma$ decreasing as $N_n$ increases
 with $\sigma \simeq 10^{-3}$ for the largest value of $N_n$.

A number of conclusions can be drawn by examining, 
for different kernels, the accuracy behavior  
 depicted in Figure \ref{fig:rho}.
The best performances are given by $M_5$ and $M_6$, which for any 
given number of neighbors $N_n$ outperform all of the other kernels. 
The cubic spline ($M_4$) and
the Wendland $C2$ ($W2$) exhibit the worst performances, regardless of the 
value of  $N_n$.

The Wendland kernels yield acceptable  density estimates  only when large 
 values of $N_n$ are used ($N_n\simgt250$), with  
$M_5$ and $M_6$ having better performances at any given $N_n$ and stable 
 estimates already for $N_n=128$.
The differences between the two families reduce progressively 
as $N_n$ is increased,  with the results becoming comparable only when 
$N_n=512$. 
\begin{figure*}
\includegraphics[height=5cm,width=16cm]{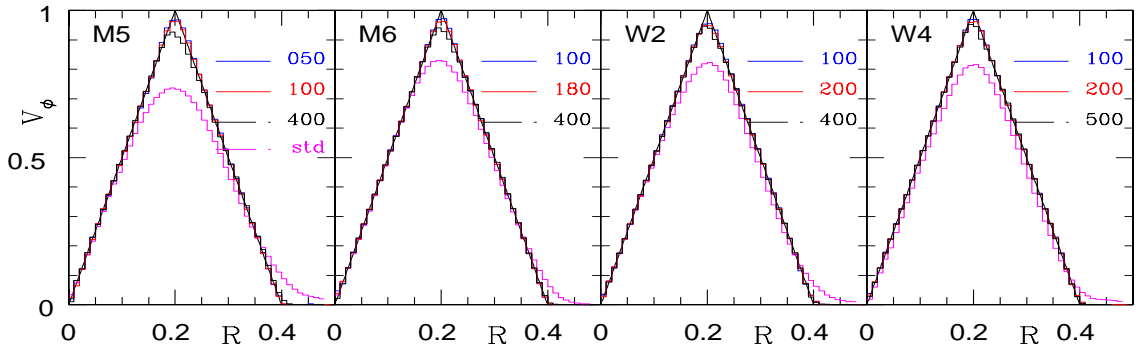}
\caption{ Azimuthal velocity profile of the Gresho-Chan vortex test 
for $M=0.34$ at $t=1$, with 1D resolution $N=128$.
Each panel is for a different kernel and, in each of them,  lines of 
different color are for different neighbor number.
For a given kernel in the standard run, the lowest neighbor number
is used.
\label{fig:vtxa}}
\end{figure*}

This behavior reflects the difference in shape between the two
families, with the Wendland kernels being more centrally peaked 
 than the B-splines and systematically overestimating the 
density. This in turn is a direct consequence of the way in which 
the Wendland kernels have been constructed in order to avoid pairing instability 
and of their spectral properties.

It must be stressed that in making comparisons between the error
behavior of kernels  of different families, it is only meaningful to compare 
 kernels with the same polynomial order \citep{ag11}.
In 3D, the Wendland equivalent of the $M_6$ kernel is therefore $W2$.

This means that in relative terms, the performances of the Wendland kernels 
are not very good unless one is willing to use a very large number of 
neighbors in the SPH simulations made with them. In this respect, it is now 
common practice 
to use  the Wendland kernels $W4$ (or even $W6$)  setting 
$N_n\simeq 200$. However, even small errors in the densities can have a 
significant impact on estimates of other hydrodynamic variables 
(see the results for the Sod  shock tube in V12). Thus, 
in the case of using Wendland kernels,  a conservative 
 lower limit on the neighbor number to be used in 3D SPH runs 
should be $N_n \simeq 400$.

{ To improve the performances of Wendland kernels   
\citet[][cf. their Eq. 18]{de12} proposed to subtract from the SPH density
estimate  (\ref{rho.eq}) a fraction of the particle $i$ self-contribution.
The correction term depends on the kernel order and neighbor number, with 
an impact which decreases as one of the two increases.
For the Wendland kernel $W4$ we show in Figure \ref{fig:rho} density 
estimated using 
the self-correction term ($W4c$), which now brings the relative density 
error down to $\sim 10^{-4}$.

However, for the hydrodynamic tests presented here the numerical set-ups 
consist of particle positions arranged in lattice or glass-like configurations
with densities of order unity.
This suggests that one can use the results of Figure \ref{fig:rho} 
to assess density errors,  which are already very small ($\sim 10^{-3}$) 
without the use of such a correction term.
Therefore for the considered runs we expect a negligible impact 
 of the self-correction term on SPH densities, and in what follows
it will not be considered.
}

These results hold for a glass-like configuration, but for a realistic 
SPH distribution of particles it is difficult to assess the error
behavior.  \citet{zu15} put the expected convergence rate between 
that found for a random distribution set ($\sigma \propto N_n^{-0.5}$) 
 and the one measured for a highly ordered distribution 
 ($\sigma \propto N_n^{-1}$), such as a glass-like configuration.
These findings strengthen the previous conclusions, suggesting that 
 when using Wendland kernels in SPH, the number of neighbors should be 
kept as high as possible.

\section{Hydrodynamic tests} \label{sec:tests}
In the following, we analyze results from some test problems aimed 
at assessing code performance of the new IA scheme.
As already outlined in the Introduction, the problems considered here have 
been chosen with the specific aim of investigating code behavior 
when subsonic flows are present in the hydrodynamic tests.

This is motivated by the serious shortcomings which affect standard SPH 
in these regimes. We first discuss the Gresho-Chan  test, which in 
this respect presents severe challenges to the SPH scheme, and then the others.

\subsection{The Gresho-Chan vortex problem} \label{subsec:gre}

The  Gresho-Chan \citep{gr90} vortex consists of a fluid of uniform density 
in differential 
rotation, with centrifugal forces balancing pressure gradients. The system
is stationary and any change in the azimuthal velocity profile that arises 
during the integration is then due to numerical artifacts.
\begin{figure}[t]
\includegraphics[height=8cm,width=8cm]{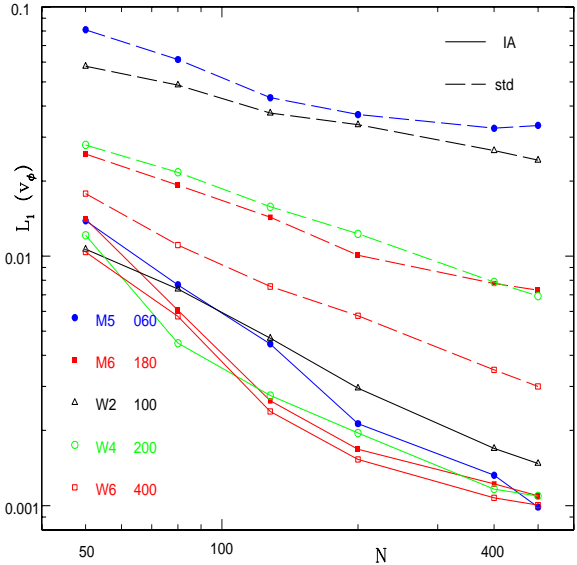}
\caption{ Convergence rate of the $L1$ velocity error for the 
Gresho-Chan vortex test. The $L1$ error is shown versus the 1D 
particle number for $M=0.34$ at $t=1$.
Dashed lines are for the standard formulation and solid lines refer
to the IA scheme.  The symbols indicate different combinations of 
kernel and neighbor number.
\label{fig:vtxL1}}
\end{figure}

Because of sampling effects, errors in force accuracy lead to noise in the
velocity field, thus generating numerical viscosity and  hence
particle disorder due to spurious transport of angular momentum.
It is then particularly problematic for standard SPH to successfully model 
this problem leaving unaltered the velocity profile during the 
simulation, and in the literature   
\citep{sp10,de12,read12,ka14,hu14,ho15,ro15,zu15}
it has been widely used to validate code performances.
 
We next describe our initial conditions set-up.
We take a gas with uniform density $\rho=1$ within the periodic domain 
$0\leq x,y <1$, with zero radial velocity, azimuthal velocity profile

\be
v_{\phi}(r)= \left\{
\begin{array}{ll}
5r  &   (0\leq r\leq 0.2) \\
2-5r  &   (0.2 < r \leq 0.4) \\
0 &   (r  >  0.4)~,
\end{array}
\right.
\label{vugr.eq}
\ee
 
and pressure profile

\be
P(r)= P_0+\left\{
\begin{array}{ll}
12.5r^2  &   (0 \leq r\leq 0.2) \\
 12.5r^2-20r+& \\
4+4\ln (5r)  & (0.2 < r \leq 0.4) \\
 2(2 \ln 2 - 1) &  (r  >  0.4)~,
\end{array}
\right.
\label{pugr.eq}
\ee
where $r=\sqrt{x^2+y^2}$, $P_0=(\gamma M^2)^{-1}$,    
 $\gamma=5/3$ and $M$ is the Mach number.
Here we have adopted the generalized  expression for the background
pressure of \citet{mic15}, so that we can consider  
 subsonic shear flows with low Mach numbers \citep{hu14}.
The standard Gresho case is recovered for $M=\sqrt{3/25}\simeq 0.34$, giving 
$P_0=5$.  

Particle positions are initialized using an $N\times N \times 16$ lattice of 
particles \citep{zu15}, $N$ being the effective 1D resolution.
For the box thickness, we set $L_z=16/N$ and we always consider  $N>32$.
 Here we use a hexagonal-close-packed (HCP) configuration for 
the particle coordinates.
The particle velocities and pressure are set according to  Equations
(\ref{vugr.eq}) and (\ref{pugr.eq}).
 All of the simulations are run up to a final time $t_f\simeq 3 M$, using a fixed 
timestep $\Delta t=t_f/(800\cdot64)$, so as to ensure the same integration 
accuracy in runs with different Mach numbers.

We quantify the convergence rate for different runs by using the $L1$ error norm 
for the velocity \citep{sp10}

\be 
L1(v_{\phi})=\frac{1}{N_b}  \sum_i^{N_b} | \overline {v}_{\phi}(i) - 
 {v}_{\phi}(r_i)|~,
\label{l1gr.eq}
\ee
where the summations is over $N_b$ bins, we set a  
 binsize of  $\Delta=0.01$  in the range $0\leq R\leq 0.5$ \citep{hu14},
$\overline {v}_{\phi}(i)$ is the average azimuthal velocity of the 
particles which lie in the $i-th$ bin interval and 
$ {v}_{\phi}(r_i)$ is the analytic solution at the bin radial 
coordinate.

Figure \ref{fig:vtxa} shows the azimuthal velocity profiles 
for $M=0.34$ at $t=1$; the one dimensional resolution is $N=128$ and 
each panel refers to a different kernel.
Within each panel, lines with different color codes are for different 
neighbor numbers, the standard run  always refers to the lowest 
neighbor number indicated in the panel.
In this case, for the B-splines,  values of $N_n$ are considered which are 
below the pairing instability threshold.

It is clear from all of the histograms that the IA scheme outperforms 
 the standard one for all of the kernels, with the latter scheme being much
more noisy. There is a tendency for the standard scheme to improve as 
higher order kernels are considered, but the error in the velocity profile
is always significant.
These results are in agreement with previous findings 
\citep{de12,read12,hu14,ro15} and clearly demonstrate how, 
for the vortex test, inaccuracies in 
gradient estimates, i.e. the $E_0$ error, are the leading error sources.
These errors are significantly reduced when using the IA scheme, showing
how good that  method  is.

To quantify the performance of the IA 
approach, for the same  test case we show in 
Figure \ref {fig:vtxL1} at $t=1$ the velocity error $L1$ as a function of 
 the 1D resolution $N$. This ranges between $N=50$ up to a 
maximum value of $N=500$.
For any given value of $N$ we considered different combinations of 
kernel and neighbor number $N_n$; these are reported in the Figure.
 For the same combination of resolution, kernel shape and neighbor number,
we performed a simulation according to the IA formulation and a 
corresponding one using standard SPH.

In the case of the B-splines we employed the highest neighbor
number that it is possible to use without having the pairing instability.
For the Gresho-Chan vortex test the convergence rate has already been
 estimated for a variety of different SPH implementations, 
  so that Figure \ref {fig:vtxL1}  can be compared with the corresponding
 rates already obtained by various authors
\citep{de12,read12,hu14,ro15,zu15}.

From Figure \ref {fig:vtxL1}  it can be seen that there is 
a resolution dependency of the $L1$ error on $N$.
In the standard case the convergence behavior is in line with 
those found previously, see for example Figure 10 of \citet{de12}
or Figure 5 of \citet{zu15}.
The dependency of $M5$ on resolution parallels that of the $W2$ kernel
and the same holds for the kernels $M6/W4$, this is very similar to what
 is seen
in Figure 10 of \citet[][we use the same neighbor number]{de12}.
However for the $L1$ norm we obtained smaller errors here. 
 For example in the $M6/W4$ case we found that for  $N=400$ 
$L1\simeq 7 \cdot 10^{-3}$, whereas for the same test run 
 their Figure 10  shows $L1\simeq 0.015$.
The same is true for the $W6$ runs, for which  here 
  $L1(N=400)\simeq 3 \cdot 10^{-3}$ is about a factor three smaller
than in their corresponding run.

When passing from the standard scheme to the IA scheme, there is a significant 
reduction in the amplitude of the $L1$ norms. 
The decrease in $L1$ is by a factor of 
 between $\sim 5$ and $\sim 10$, with some dependency on resolution and 
adopted kernels. For the $M5~(N_n=60)$ kernel the ratio between the norms 
ranges from $L1(IA)/L1(std)\simeq 1/8$ at $N=50$ down to $\sim 1/30$ 
when $N_n=500$.

We now analyze the convergence rate of $L1$  in the IA formulation. 
This depends on the adopted kernel, and for $M5$ we found 
$L1 \propto N^{-1.2}$. This is close to what was given by \citet{sp10} 
when using the moving mesh code Arepo ($L1 \propto N^{-1.4}$).

To estimate the rate for the other kernels we adopt a conservative 
view and only include in the fit those points with $N\geq80$.
We then obtain $L1 \propto N^{-1}$, which is better than that reported
by \citet[][$L1 \propto N^{-0.7}$]{hu14}
 for their pressure-entropy SPH formulation. 
The rate is also in agreement with that shown by 
\citet[][Figure 10, case $F3$: $L1 \propto N^{-1}$]{ro15}, 
who implemented  a IA scheme using a $W6$ kernel. Note, however, that the 
value of $L1$ at $N=300$ found there is a factor $\sim3$ higher than that 
found here ($L1 \sim 10^{-3}$).

Higher convergence rates have been obtained by 
\citet[][$L1 \propto N^{-1.4}$]{read12}, who employed a modified version of SPH 
with high-order dissipation switches, and by \citet{zu15}. The latter investigated
the convergence behaviour  of  standard SPH showing that 
consistency in numerical convergence is achieved when the conditions  
$N\rightarrow \infty ~h \rightarrow 0 ~N_n \rightarrow \infty$ are 
satisfied. In their varying $N_n$ case ($N_n=120\cdot (N/32)^{1.2})$,
for the Gresho-Chan vortex  test the authors report 
$L1 \propto N^{-1.2}$, a  much faster rate than that obtained by 
keeping $N_n$ constant. The value of $L1$ in the $N=500$ case is 
of the same order ($L1 \simeq 3 \cdot 10^{-3}$) as that obtained here 
for the  $W6$ standard run with the same resolution.
\begin{figure}[t]
\includegraphics[height=8cm,width=8cm]{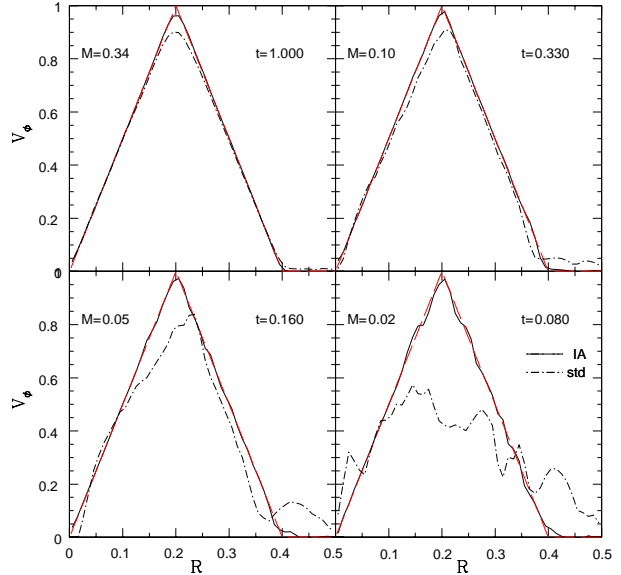}
\caption{ Velocity profiles of the Gresho-Chan vortex test 
for $M=0.34,~0.1,~0.02,~0.05$ (clockwise from top-left)
at the final times $t_f\simeq 3 M$. The tests have been performed 
using $N=128$ and the $M6$ kernel with $N_n=180$ neighbors.
Solid lines are for the IA formulation , dot-dashed lines refer
to the standard scheme. The dashed line in color is the analytical 
solution.
The Figure can be compared with Figure 2 of \citet{hu14}.
\label{fig:vtxlow}}
\end{figure}

Finally, with the exception of the lowest order kernels ($M5$ and $W2$), 
 a comparison of the $L1$ norms 
with those produced using the moving-mesh code Arepo shows that 
the IA formulation gives results which are comparable or  better than
those obtained with the mesh code \citep[][Figure 29]{sp10}, with 
 $L1$ ranging here from $L1 \simeq 8\cdot 10^{-3}$~($N=80$) down to 
$L1 \simeq 10^{-3}$~($N=500$).

To further investigate the performance of the IA scheme, we ran 
a suite of vortex tests with progressively lower Mach numbers.
These tests are particularly challenging since, at constant velocity, the lower
 is the Mach number, the higher is the sound speed. This in turn implies an 
increase in the viscous force. Errors in the momentum equation 
then become progressively more important.
 
To aid comparison with the previous works, as in \citet{hu14} we 
considered the following Mach numbers $M=0.02,~0.05,~0.1$ and $M=0.34$.
We ran the simulations using $N=128$ as 1D resolution. 
This is a factor of two lower than that used by \citet{hu14}
in their tests, however the results are not significantly 
affected by this choice.
For the $M6$ kernel ($N_n=180$) we show the azimuthal velocity 
profiles of the four test cases in Figure \ref{fig:vtxlow}, so that 
the Figure can be compared with the corresponding histograms of 
Figure 2 of \citet{hu14}.
\begin{figure}[!t]
\includegraphics[height=11cm,width=8cm]{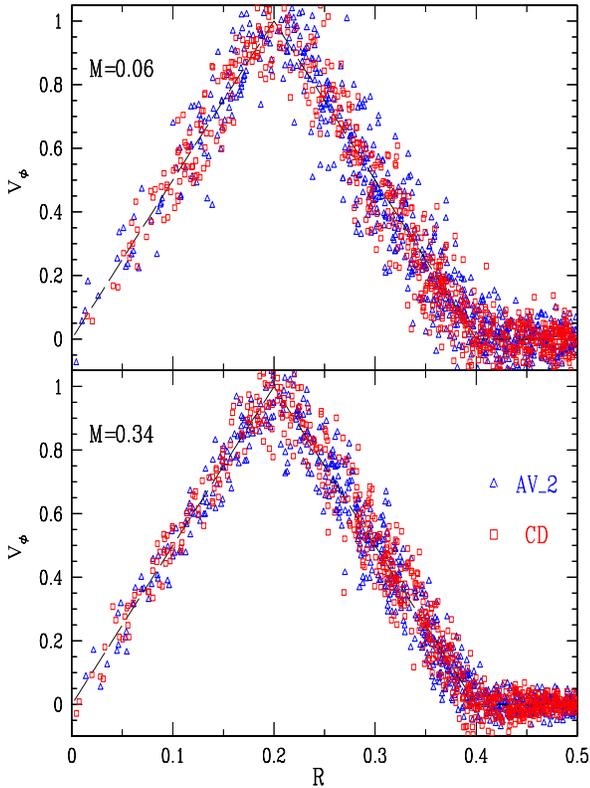}
\caption{ 
Plots at $t=3$ of the azimuthal particle velocities for two Gresho-Chan 
vortex tests, both performed using a one-dimensional resolution of 
 $N=128$ and the $M6$ kernel with $N_n=180$ neighbors.
Bottom panel is for $M=0.34$ and top panel for $M=0.06$ with 
 $P_0=50$.
The two plots  can be compared with the corresponding ones in Figure 4 \& 5 
 of \citet{ho15}.
For the sake of clarity we show velocities of a subset of randomly 
selected particles.
Open triangles are for the time-dependent AV scheme with settings AV$_2$ 
(See Section \ref{subsec:visco}), open squares refer to the AV switch of 
\citet{cul10}.
\label{fig:vtnoise}}
\end{figure}

The profiles of the standard runs largely reproduce those 
of \citet{hu14}; however, a striking feature of the IA scheme 
which emerges from the histograms of Figure \ref{fig:vtxlow} is the close 
proximity of the azimuthal velocity profiles to the  analytical 
solution.
This occurs even when very low Mach numbers are considered, as can 
be seen from the $M=0.02$ case.
This shows  the effectiveness of the IA method for eliminating sampling errors in 
SPH when subsonic flows are present.

Moreover, these findings are in agreement with previous results 
\citep{read12,hu14}  and
demonstrate that in SPH simulations of the Gresho-Chan  test,
errors in force accuracy dominate over viscous effects.

{
For the considered tests we have shown until now the mean binned velocities.
In order to assess the amount of noise present in the various runs, it is 
useful to plot directly the azimuthal particle velocities. 
To this end we ran two tests with  
 1D resolution $N=128$, Mach numbers  $M=0.34$ and $M=0.06$, respectively.
In the latter case we set a background pressure of $P_0=50$.
For each run performed using the time dependent AV scheme  with settings AV$_2$
( see Section \ref{subsec:visco} ), we also consider a parent simulation in
 which has been implemented  the AV switch of  \citet{cul10}.

The results are shown in Figure \ref{fig:vtnoise} at $t=3$, where for the 
two test cases we plot the azimuthal velocities for a subset of all 
particles. The velocity distributions can be compared directly 
with those of the corresponding runs in Figure 4 \& 5 of \citet{ho15}.
 An important feature which emerges from the plots of Figure \ref{fig:vtnoise} is 
that both AV methods show velocity distributions which are evenly scattered 
around the analytic solution, with the AV switch of \citet{cul10} exhibiting
 a much smaller  amount of noise.
It is worth noting how the IA method, even for very low Mach numbers, 
 can accurately follow the analytic solution also 
at the peak vortex velocity.
This   behaviour is much better than that  seen for the same test in the top 
panel of Figure 5 of \citet{ho15}.

Finally, it must be pointed out that in these simulations  the amount of thermal 
diffusion due to the AC term 
is negligible and  we do not include such a term in the SPH equations.
This occurs because of the high sound speeds in the low Mach number regime, 
so that the time evolution (\ref{alfac.eq}) of the $ \alpha^C_{i} $ parameter
is driven by  the decaying rate $1/\tau^C_i$, which  dominates over the source 
term $S^C_i$.
To better quantify this issue we use Equation (\ref{dudis.eq}) to estimate at 
time $t$ the 
change in thermal energy due to the AC term:  $\Delta u_{AC}  \simeq  t ( D^{AC} \nabla^2 u ) $.
\begin{figure*}[!t]
\includegraphics[height=11.2cm,width=17.2cm]{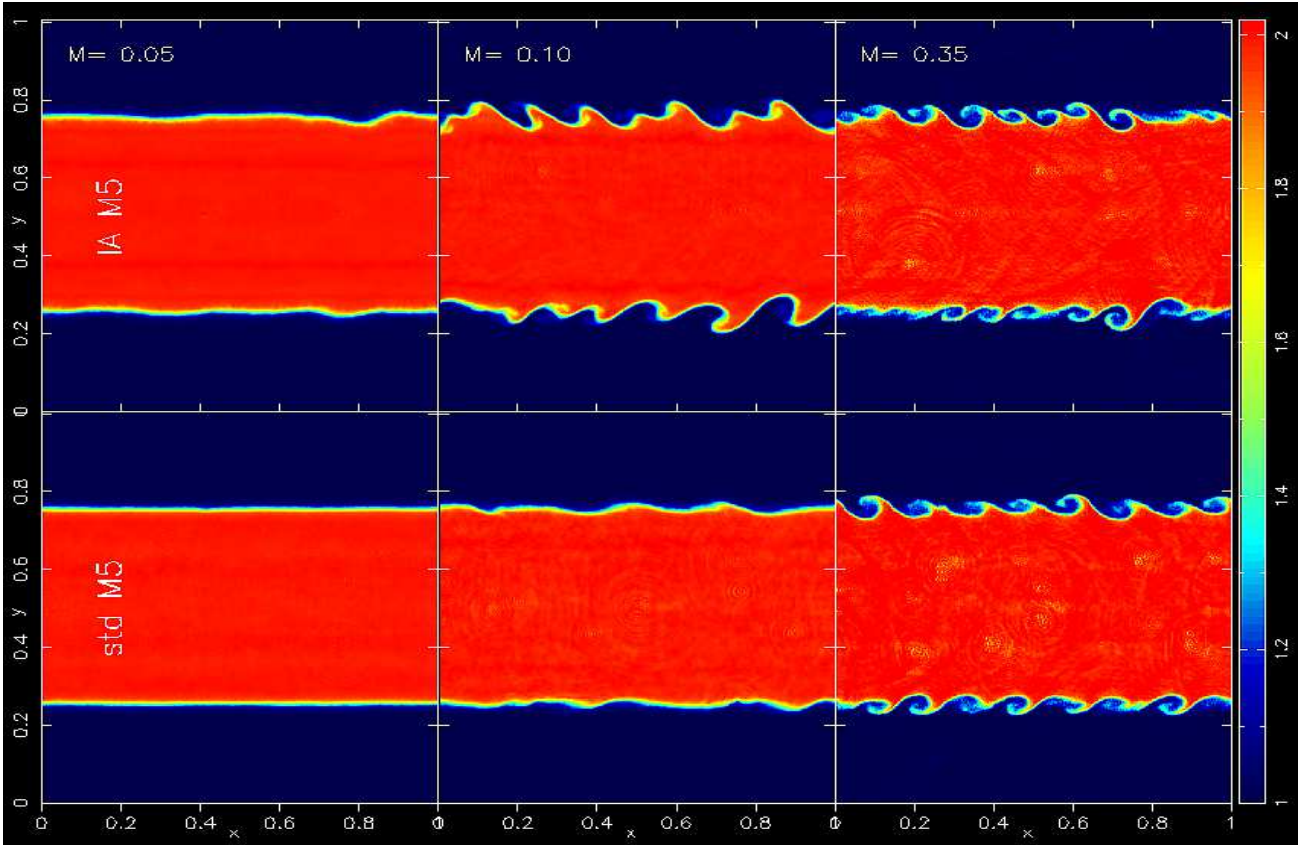}
\caption{Density maps at $t=\tau_{KH}$ for the 2D KH instability
tests described in Section \ref{subsec:kh}. 
The tests have been performed with a density contrast of
$\chi=2$ between the two contact layers.
From  left to right the
different panels are for different Mach numbers: $M=0.05,~M=0.1$ and $M=0.35$.
Each test case was run separately with  both the
standard (bottom) and the IA (top) scheme. All of the maps have been
 extracted from simulations performed using the $M5$ kernel.
\label{fig:khmap}}
\end{figure*}

The Laplacian of $u$ has a maximum at $r=0.2$, 
where the azimuthal velocity reaches its peak value.
The diffusion coefficient  is then  given by 
$D^{AC}_i\simeq \alpha^C_{i} 5 r_{ij} r_{ij}/2\simeq 5  \alpha^C_{i}h_i^2/2$, 
so that

\begin {equation}
\Delta u_{AC}  \simeq  \frac{5}{2}   \alpha^C_{i} 
 h_i^2 \nabla^2 u_i t = \frac{5} { 2} 
 (S^C_i \tau^C_i )  h_i^2  \nabla^2 u_i t~,
\end {equation}

where we have approximated $\alpha^C_{i}$ with the equilibrium solution
$ \alpha^C_{i}(eq) \simeq S^C_i \tau^C_i \simeq h_i^2  \nabla^2 u_i/ 
(0.2 c_i  \sqrt{u_i} )$.

For $M=0.34$  at $r=0.2$   $ u \sim 15/2, \nabla^2 u \sim 300/4$,
$c_i \sim 3$   and $h_i$ can be easily 
computed because $\rho=1$ so that 

\begin {equation}
\Delta u_{AC}  \simeq 8.6 \cdot 10^3  h_i^4  t   
\end {equation}

For  $N\simgt50$ and $ t\simlt3$ this term is always much smaller than $u$, 
regardless of the chosen kernel.
}

\subsection{The Kelvin-Helmholtz instability} \label{subsec:kh}

The KH instability has been investigated by many authors since it 
is a classic test in which SPH fails to properly model the 
development of the instability \citep{pr08,read10,vrd10,ju10,hs10,ch10,
mu11,va12,na12,ka14,hu14,ho15}.

The test consists of two fluid layers of different densities 
sliding past each other with opposite shearing velocities,   
 and a small velocity perturbation is imposed 
in the direction perpendicular to the contact surface. A fluid
instability develops, which is initially small and then becomes progressively
larger until non-linearity is reached with the appearence of 
KH rolls. 
For a sinusoidal perturbation of wavelength $\lambda$, a linear 
time scale can be defined as
\be
\tau_{KH}=\frac{\lambda (\rho_1+\rho_2)}
{\left(\rho_1\rho_2\right)^{1/2}v}~,
\label{taukh.eq}
\ee
where $\rho_1$ and $\rho_2$ are the two fluid densities with a density
ratio $\chi=\rho_1/\rho_2$ and $v=v_1-v_2$ is the
relative shear velocity.
\begin{figure*}[!t]
\includegraphics[height=6.4cm,width=17.2cm]{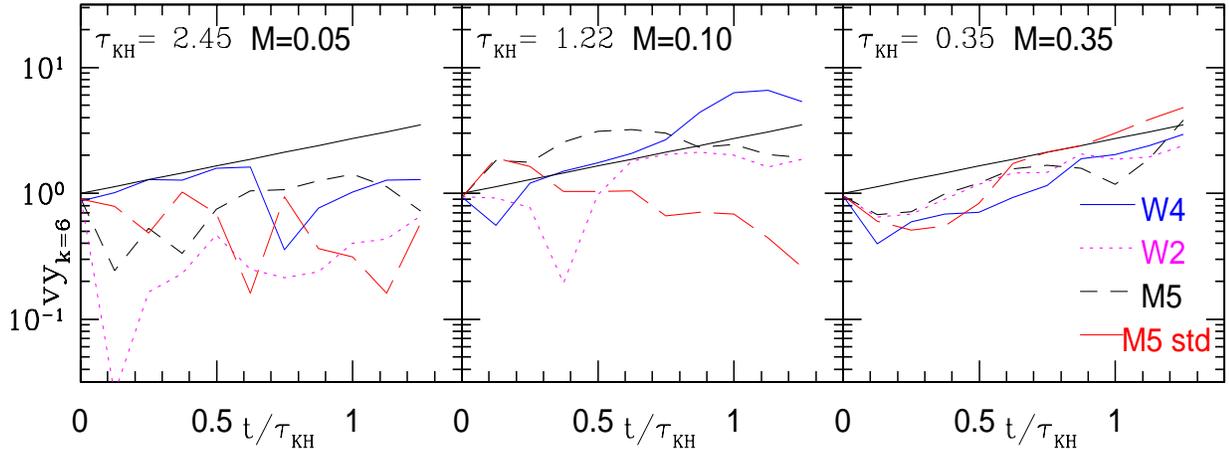}
\caption{Growth rate of the $\lambda=1/6$ velocity amplitude 
as measured by making a Fourier transform of $v_y$. 
The KH tests  are for a density contrast $\chi=2$ and
 three different values of Mach number have been considered
($M=0.05,0.1,0.35$). For each KH test case we ran IA simulations 
using the $M_5$, $W2$ and the $W4$ kernels.
Additionally, for the $M_5$ runs, we also performed a corresponding 
standard simulation. The solid line is the linear theory growth rate expectation
$\propto e^{t/\tau_{KH}}$, normalized to the numerical amplitude at $t=0$.
\label{fig:khtau}}
\end{figure*}

To perform the test, the following conditions   

\be
\rho,~T,v_x=\left\{
 \begin{array}{ l l }
\rho_1,T_1,v_1  & |y-0.5|\leq 0.25 \\
\rho_2,T_2,v_2  & |y-0.5| >  0.25
 \end{array}
\right.
\label{rhokh.eq}
\ee
are applied for a fluid with adiabatic index $\gamma=5/3$ 
in a two-dimensional periodic domain with
cartesian coordinates $x\in \{ 0,1 \}$, $y\in \{ 0,1 \}$. 
We set here  $\rho_2=1$, and $\chi=2$  for the density contrast.  

The two layers are in pressure equilibrium with $P_1=P_2=5/2$, so that the
 sound velocities in the two layers are 
$c_2=\sqrt{{\gamma P_2}/{\rho_2}}=2.04$ and
  $c_1=c_2/\sqrt{\chi}=1.44$, respectively.  
The Mach number of the  high-density layer is 
$M\simeq v_1/c_1 \simeq 0.7 v_1 $ and the KH time scale is
$\tau_{KH}\simeq 0.177 /v_1$. 
We ran KH simulations with three different Mach numbers: 
$M=0.05,~0.1$ and $M=0.35$. For the latter value the initial condition set-up
was similar to that of \citet[][Section 4.4.1]{ho15}.

 The KH instability is triggered by adding in the proximity of the layer 
boundaries a small single-mode velocity perturbation  along the $y-$direction

\be
v_{y}=\delta v^{(0)}_y \sin(2\pi x/\lambda)~,
\label{eq:vy}
\ee
where   $\delta v_y^{(0)}=2\cdot10^{-2}v_1$, $\lambda=1/6$ and $v_y=0$ 
 if $|y-\sigma|>0.025$, where $\sigma$ takes the 
values $0.25$ and $0.75$, respectively.
Note that for the amplitude of the initial velocity perturbation 
$\delta v^{(0)}_y$
we set here, 
unlike in previous runs (V12),  a relative 
constant amplitude with respect to the streaming velocity.
{
This was done in order to consistently compare,
 between runs with different Mach numbers, the impact of
 zeroth-order errors on  $v_y \propto M$.
}

We performed  the initial condition set-up by arranging $N^2=512^2$ equal mass 
particles inside the simulation box, 
setting the particle coordinates according to an  HCP configuration.  
The lattice spacing was smoothly adjusted at the fluid interfaces so 
as to avoid density discontinuities; the details of the whole procedure 
are given in V12.
For each  test case the IA simulations were then performed 
using the $M_5$, $W2$ and the $W4$ kernels with 
neighbor number $N_n=50$, $N_n=72$ and $N_n=162$, respectively. 
For the $M5$ runs we also considered standard simulations.
Finally, all of the simulations were performed with the AC term of 
Section \ref{subsec:ac} switched on.

For the specified  range of Mach numbers, we first
show in Figure \ref{fig:khmap} density plots of the KH simulations at 
$t=\tau_{KH}$. We show  maps extracted from the $M5$ runs, and 
 contrast the IA scheme against the standard SPH scheme.
For $M=0.35$,  both of the methods  are able to produce 
KH rolls. At lower Mach numbers ($M=0.1$) the standard method 
completely fails the KH test, whereas the IA scheme shows a degraded capability 
to resolve KH rolls.
At $M=0.05$  the rolls are absent and the differences between 
the two schemes are  no longer present.

The relative performances of the two methods can be quantitatively assessed 
by measuring the $E_0$ error \citep[][V12]{read10,vrd10} for the various runs.
{ Here we first show }the growth rate of the KH instability 
\citep[][V12]{ju10,hs10}, 
which allows one to recognize in a more visual way the 
differences between the KH results produced by the two schemes.
The growth rate is measured by Fourier transforming, at different times, 
the $\lambda=1/6$  growing mode of the  $v_y$ velocity  perturbation 
\citep[][V12]{ju10}. 

The growth rates are shown in Figure \ref{fig:khtau}, and their relative 
differences
confirm the visual impressions  derived from the maps of Figure \ref{fig:khmap}.
For $M=0.35$ there are no significant differences between the two methods 
and both are able to follow the growth of the KH instability
 (Figure \ref{fig:khmap}, right panel), the rates of the different runs
being in accord with the analytic expectation.
The results can also be compared with the corresponding rates 
in Figure 18 of \citet{ho15}, taking care about the different time scales 
due to the different number of modes used to seed the perturbation.
Note that, unlike in \citet{ho15}, the standard version here correctly follows
 the development of the KH instability. We interpret this
 difference as being due to the small neighbor number ($\sim32$ in 3D) 
adopted in his standard (PSPH) run.
\begin{figure}[t]
\includegraphics[height=8cm,width=8cm]{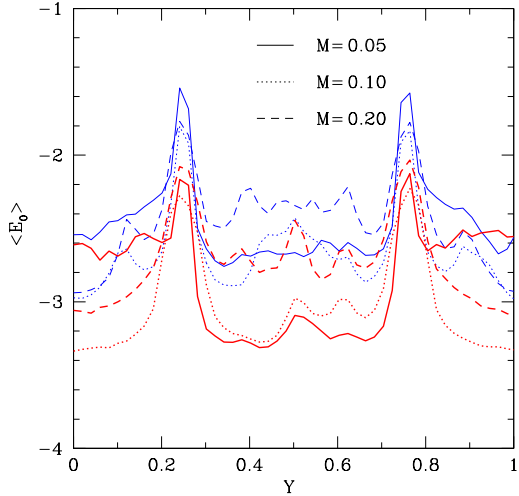}
\caption{  Averaged binned distribution of the particle errors $|\vec E_i^0 |$  
versus $y$ for the KH runs of Figure \ref{fig:khmap}. 
Different lines are for different Mach numbers. Thick red (thin blue) 
lines refer to IA (standard) runs.
\label{fig:kherr}}
\end{figure}

As lower Mach numbers are considered, from the other panels of 
Figure \ref{fig:khtau} one can see a growing 
difficulty of the IA scheme in following the KH instability, regardless of the 
kernel employed in the simulation.
This happens because by reducing the Mach number the shear velocity is also
reduced and in turn, owing to the chosen settings, the initial velocity 
amplitude is reduced as well. At a fixed resolution the impact of gradient 
errors, and the 
subsequent particle disorder, on the growth of the KH instability is then 
higher as the Mach number decreases.

{
The $ \vec E_0 $ error of particle $i$   is defined by
 \citep{read10}

 \be
    \vec {E_i}^{0}=\sum_j \frac{m_j}{\rho_j} \left[
  \frac{\rho_i}{\rho_j} +\frac{\rho_j}{\rho_i}\right]
   h_i \vec \nabla_i \bar W_{ij}~,
  \label{en0.eq}
  \ee
and  we show in Figure \ref{fig:kherr} the mean binned 
distribution of the  particle errors  versus $y$. The plots refer
to the runs of Figure \ref{fig:khmap}.
A key feature is the magnitude of the errors, which 
in  proximity of the interfaces 
for the $IA$ runs are smaller by  factor $\sim 5$ than
the standard ones.
This is in line with what expected and in accord with 
what seen in Figure \ref{fig:khtau},  
with the growth rates of the IA runs exhibiting a better 
behavior at  low $M$.
However  one can see from Figure \ref{fig:khtau} that for $M=0.05$ the KH
instability is not correctly reproduced even in the IA scheme.
In such a case gradient errors can be reduced by increasing
the simulation resolution.
}

We do not undertake here a resolution study aimed at assessing 
the convergence rate to the KH solution in the very low ($M\simlt0.1$) 
subsonic regime.
We use instead a simple argument to provide a rough estimate of the 
minimum number of particles $N^2$ which would be necessary to 
 simulate the $M=0.05$ KH test case.

An $L1$ error norm for the KH problem has been introduced by \citet{rb10} 
and, in analogy with their Equation 11, we conjecture here for $L1$ a 
generic dependence of the form 

\be
L1 \propto N^{-2\alpha} (1+t)^{\gamma(N)}~,
\label{eq:l1}
\ee

on the particle number $N_p=N^2$, and the simulation time $t$. 
We have dropped the dependency on the bulk flow velocity, present 
in their Equation 11, and 
for the  power-law dependencies we generically assume   
the  exponents $\alpha$ and $\gamma(N)$.

The simulations of \citet{rb10} were 
performed using the  Eulerian mesh code ART \citep{kr97}; note however that 
their  initial condition setup
corresponds here to the $M=0.35$ KH test case.

The ratio between  the error norms of two different KH runs 
 is then 

\be
L_2/L_1 = (N_1/N_2)^{2\alpha}  
(1+t_2)^{\gamma_2}/(1+t_1)^{\gamma_1}~,
\label{eq:l2}
\ee

where we set $t=\tau_{KH}\simeq0.124/M$ in order to 
 consistently  compare the norms and $\gamma_i\equiv \gamma(N_i)$.
We now assume as reference run the $M=0.35=M_1$ test case, for which 
from Figure \ref{fig:khtau}   $N=512$ can be considered an adequate resolution 
up to $t=t_1\sim0.35$. 
Therefore for $M_2=0.05=M_1/7$ the norm ratio is 

\be
L_2/L_1 = (512/N_2)^{2\alpha}  
 (3.45)^{\gamma_2}/(1.35)^{\gamma_1}~.
\label{eq:l3}
\ee

For the dependency on simulation resolution \citet{rb10} 
  report  $\alpha=1$ for their Eulerian code.
Here the numerical convergence is likely to be shallower, with $\alpha<1$.
However, the results of Section. \ref{subsec:gre} indicate for the vortex
test a convergence rate of the IA scheme very close to that
seen using moving mesh schemes \citep{sp10}.
We therefore assume here $\alpha=1$ as a reasonable slope on  
resolution convergence, thus putting a conservative lower limit 
on $N_2$.

In \citet{rb10}, the time evolution of the error norm  has a weak 
dependency on numerical resolution: $\gamma(N)\simeq 2(64/N)^{0.5}$.
This slope  clearly depends on the adopted numerical scheme and 
we simplify this dependency by assuming $\gamma_i=1$. 
The impact of this assumption on estimating $N_2$ is however 
relatively unimportant, the ratio between the two time factors
 being in any case of order unity and closer to one as $\gamma_i<1$.
In fact, we further simplify the ratio (\ref{eq:l3})  by just 
removing the time factors.

Finally, an accuracy criterion for the simulations is set by putting 
an upper limit on the error norm $L1 \simlt   err(L1)$,
with $err(L1)$ being a given threshold.
We now assume for $err(L1)$ a generic dependence on the 
initial velocity amplitude $\delta v^{(0)}_y$ of the form 
$err(L1)\propto (\delta v^{(0)}_y)^{\beta}$, with $\beta\geq1$.
This lower limit on $\beta$ is justified by the requirement that 
 lower values of 
 $\delta v^{(0)}_y$  must correspond to lower values of $err(L1)$.
Then , for the ratio (\ref{eq:l3}), we have

  \begin{eqnarray*}
\lefteqn{L_2/L_1 } &  & ~~~
 =(512/N_2)^{2} \simlt
\left( \frac{ \delta v^{(0)}_y(2)}{\delta v^{(0)}_y(1)}\right)^{\beta}\\
 &  & = (M_2/M_1)^{\beta}=(1/7)^{\beta}~. 
    \label{eq:l6}
   \end{eqnarray*}

Setting $\beta=1$ we thus obtain for $N_2$ the lower limit 
$N_2\simgt 1500$.
Note however that to achieve numerical convergence in a 
consistent way in SPH the number of neighbors $N_n$ must also increase when
$N\rightarrow \infty$ and $ h \rightarrow 0$ \citep{zu15}.
For their Gresho-Chan vortex  test  \citet{zu15} 
adopt,  when $N_n$ is allowed to vary, $N_n\propto N_p^{1.2}$. 
In such a case, by referring to the $M=0.35 ~W4$ run with
$N_n=162$ neighbors, we conclude that $N_n\simgt7\cdot162=1134$ 
neighbors and $N_p\simeq 2\cdot 10^6$ particles 
are the least necessary in order to simulate the $M=0.05$ KH test case using 
the $W4$ Wendland kernel.

{
Finally, it is worth noting that the difficulties of SPH to follow the 
formation of 
KH instabilities  depend not only on velocity noise, but also on the 
 local mixing instability  \citep[LMI:][]{ag07,pr08,read10,vrd10}.
This LMI occurs because, in the presence of a density step, the entropy
conservation of SPH causes a pressure blip at the boundary.
These pressure discontinuities in turn lead to the presence of shock waves 
which then inhibit the growth of KH instabilities \citep{vrd10}.

Different approaches have been taken to eliminate or reduce the LMI: 
by introducing initial conditions with a smoothing of the density step 
\citep{vrd10}, and/or 
adding an AC term to give smooth entropies \citep{pr08}, or by 
reformulating the SPH density estimate \citep{ri01}.

Based on a suite of numerical tests, \citet{vrd10} argued that 
for low Mach numbers ($M\leq0.2$) the growth of KH instabilities 
is still suppressed by the LMI. Although the magnitude of the shocks 
induced by LMI has been greatly reduced because of the initial 
density smoothing, \citet{vrd10} found that for low M the 
time scales $\tau_{KH}$ are much higher  than those set by the numerical shocks.

It is not trivial to remove from the simulations these residual 
shocks.  
For instance, they can be eliminated by applying a relaxing scheme to 
the initial conditions, but the growing KH instabilities 
are then strongly suppressed by the induced particle disorder \citep{vrd10}.
A study on these effects is beyond the scope of this paper.
}

{
As a final point, it must be stressed that the results of the
KH runs presented here  have been obtained by using the AV scheme of 
Section \ref{subsec:visco} with settings AV$_2$.
By replacing this scheme with the AV  switch of  \citet{cul10} 
we expect a significant reduction in the amount of AV 
present in the simulations ( see results of the previous and next 
Section ).
This in turn will result in  a more inviscid behavior and a better capacity
of the code to follow the development of the KH instabilities. 
}

\subsection{Subsonic turbulence} \label{subsec:subs}

Studies of driven isothermal subsonic turbulence  
\citep{ba12} have shown substantial differences in the properties 
 of the velocity power spectra extracted from mesh-based 
simulations, when compared with those produced from the 
same test runs using the standard formulation of SPH.

Although the use in standard SPH of a time-dependent AV
scheme alleviates the problem \citep{pr12b}, the discrepancies are
still present and their origin has been identified 
as being due to large errors in the SPH gradient estimates \citep{ba12}.
These errors in turn imply the presence of subsonic velocity
noise which is higher as lower Mach numbers are considered.
As a result, SPH simulations exhibit spectra with a much 
smaller inertial range (i.e. Kolgomorov-like) than the ones 
measured using mesh codes.

A faithful numerical modeling of subsonic turbulence is 
particularly relevant in various astrophysical contexts
(star formation, intracluster medium, intergalactic medium), and
it is therefore important to investigate the capability of the IA
scheme to properly simulate this test problem.

To this end we set an HCP lattice of $N^3$ particles with initially 
zero velocities inside a 
periodic box of sidelength $L=1$ and density $\rho=1$.
The gas was isothermal with $\gamma=1$ and $c_s=1$.
Turbulence in the gas was driven by adding to the momentum Equation
(\ref{fsph.eq}) of the particles an external stochastic 
driving force $\vec {a}_{stir}$.
This was constructed in $k$-space according to a procedure already 
used by previous authors \citep{pr10,ba12,pr12b,ho13,ho15,zu15}

The power spectrum of $\vec {a(k)}_{stir}$ varies as 
  $P(k)\propto k^{-5/3}$ and the Fourier modes are non-zero in the 
range between $k_{min}=2 \pi/L$ and $k_{max}=2 k_{min}$.
The phases of the stirring field are drawn from an Ornstein-Uhlenbeck
(UO) process for which  the random sequence at the step $n$ is given by 
\citep{es88,ba01}

\be 
x_{n+1}=f x_n +\sigma \sqrt{1-f^2} z_n~,
\label{xnsub.eq}
\ee

where $f=\exp{\left(-dt/t_s\right)}$ is a decaying factor, 
 $z_n$ is a Gaussian random variable  with unit variance 
and $\sigma$ is the variance of the UO process. The constructed sequence then has 
$<x_n>=0$ and  $<x_{n+1} x_n>=\sigma^2 f$ .

In order to obtain a pure solenoidal driving, we apply an Helmholtz 
decomposition in $k$ space:

\be
a_i(k,t)=b_i(k,t)-k_i (\vec b \cdot \vec k)/k^2~,
\label{afsub.eq}
\ee

where the vector $\vec b (\vec k)$ is a complex 
vector-valued stochastic process characterized at any given $\vec k $  
by 6 UO  random sequences 
(\ref{xnsub.eq})  and $\vec a$ is the solenoidal stirring field 
($\vec a \cdot \vec k=0$).

The particle accelerations are calculated at each timestep
by updating the stochastic field according to the described procedure, 
 the summation in $k$ space being performed  by summing 
 directly at the particle positions. 
For the driving parameters we use the values of \citet[][Table 1]{ba12}.
The power spectrum is normalized so that the rms Mach number 
${\mathcal M}$  lies 
in the range ${\mathcal M}\sim 0.25-0.3$ after the simulations have 
reached the steady-state regime ($t\simgt 5$).
\begin{figure}[t]
\includegraphics[height=8cm,width=8cm]{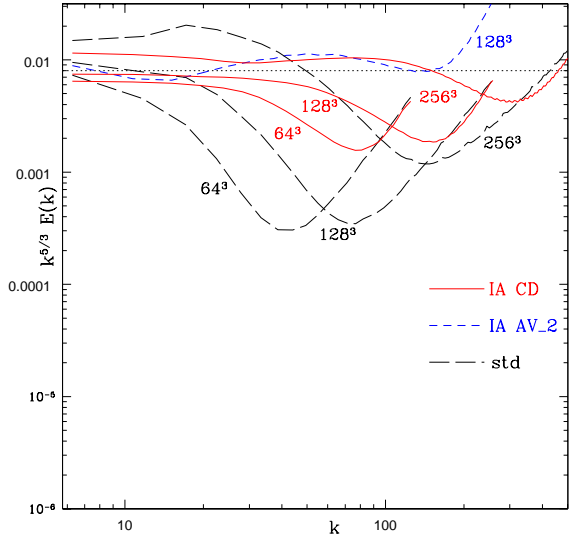}
\caption{ Time-averaged velocity power spectra of driven subsonic 
(${\mathcal M\sim0.3}$) isothermal turbulence.
The spectra are compensated by $k^{5/3}$ so that the horizontal dotted 
line indicates the Kolgomorov scaling. 
We ran simulations using the same driving routine with 
 $N^3=64^3,~128^3$ and $N=256^3$. 
Dashed (black) lines are for the standard SPH runs, 
short-dash (blue) line is the $N^3=128^3$  IA run 
with AV settings AV$_2$ (Section \ref{subsec:visco}), solid (red) 
lines are  the  IA runs performed using the 
 the AV switch of \citet{cul10}. 
\label{fig:pwsubs}}
\end{figure}

We compared results extracted from subsonic simulations performed with
the standard and IA implementations of SPH. We ran simulations with 
three different resolutions: $N^3=64^3,~128^3$ and $N=256^3$. 
For a given resolution we used in both of the schemes the same initial 
condition set-up and stirring force field.
In all of the simulations we ran up to $t=50$ and adopted the $M5$ kernel 
with $N_n=50$ neighbors.
{We perform standard and IA runs by using the time-dependent AV scheme
of Section \ref{subsec:visco} with settings AV$_2$. Additionally, we also 
run a set of IA simulations by using the AV method of 
  \citet{cul10}. 
In the following, we will refer to these IA runs with the term IA-CD, whilst 
we will use the term IA-AV when referring to  the IA runs with AV settings 
 AV$_2$.  
}

As in other works \citep{ba12,ho15} we measure the  spectral properties of the 
turbulent velocity field  to assess the performances of the two codes.
The velocity power spectrum  is defined as
   \begin{equation}
E(k)=2 \pi k^2 \mathcal{P}(k)~, 
  \label{ekasub.eq}
 \end{equation}

 where $k\equiv|\vec k|$, 
 and  $\mathcal{P}(k)$  is the ensemble average velocity power spectrum.
  This is given by
   \begin{equation}
< \vec {\tilde u}^{\dagger}(\vec k^{\prime}) \vec {\tilde u}(\vec k)>=
\delta_D(\vec k^{\prime}-\vec k) \mathcal{P}(k)~,
    \label{pwftsub.eq}
   \end{equation}
 
where $ \vec {\tilde u}(\vec k)$ is the Fourier transform of the velocity
field $\vec u (\vec x)$:

   \begin{equation}
 \vec {\tilde u}(\vec k)= \frac{1}{(2\pi)^3} \int \vec u (\vec x) 
 e^{-\imath 2 \pi \vec k \cdot \vec x}d^3x~.
    \label{vkftsub.eq}
   \end{equation}

 In the case of incompressible turbulence, the energy spectrum follows
 the Kolgomorov scaling $E(k) \propto k^{-5/3}$.
To measure the energy spectrum we first set inside the simulation box 
a cube with $N_g=(2N)^3$ grid points. 
From the particle velocities $\vec u(\vec x_i)$
we then estimate  the grid velocity field $\vec u(\vec x_g)$ at the grid 
points $\vec x_g$, using a 
triangular-shaped cloud function (TSC) interpolation scheme.

We then compute the discrete Fourier transforms
 of $\vec u(\vec x_g)$ 
 and  the discrete power spectrum
$\mathcal{P}^d(k)=< |\vec {\tilde {u}}^d(\vec k)|^2>$
is evaluated by binning the quantity
$|\vec {\tilde u}^d(\vec k)|^2$ in spherical shells of radius
$k$ and averaging in the bins.
The energy density of Equation (\ref{ekasub.eq}) is then given, 
 aside from a normalization factor, 
by $E(k)=2 \pi k^2 \mathcal{P}^d(k)$.
Finally, for a given simulation, the spectrum $E(k)$ is estimated 
by doing a  time-average between $t=10$ and $t=25$, with the spectrum
being sampled each $\Delta t=0.08$ time interval.

We show in Figure \ref{fig:pwsubs} the spectra $E(k)$ as measured from 
 our simulations.
The spectral behavior of the standard runs is in broad agreement with 
previous findings \citep{pr12b,ba12,ho13,ho15,zu15}. The spectra are characterized 
by a very narrow inertial range at low wavenumbers, with a significant 
decline at higher $k$. The spectra reach a minimum at 
a wavenumber $k_{turn}$, which increases as higher resolutions are 
considered, followed by a steep increase in the power at smaller scales 
$k\simgt k_{turn}$.

The precise value of $k_{turn} \sim 40-200$ depends on $N$, but it is still 
much smaller 
than the noise scale $\sim 2h_{max}$ set by the kernel. 
 For incompressible turbulence one can easily approximate $h_{max}$ 
with the value of $\overline h$ given by the average density, thus 
obtaining 
$ 2 \pi / 2 {\overline h} \sim  \pi \zeta (N /L ) 
( 4 \pi / 3 N_n )^{1/3} \sim 220 (N/64) >> k_{turn}(N)$

According to \citet{pr12b} the very limited capability 
of standard SPH to develop a Kolgomorov-like spectrum is due to 
the excess of numerical viscosity present in the scheme,  which can 
be reduced by adopting a time-dependent AV switch. 
\footnote{ We recall that  with the term standard SPH 
we refer here to the usual SPH scheme of Section \ref{subsec:method}, but 
incorporating the time-dependent AV switch  
described in  Section \ref{subsec:visco} }. 
In contrast, \citet[][Figure 6]{ba12} showed that the rise in power at small 
scales is mainly a result of the subsonic velocity noise due to
kernel gradient errors present in standard SPH.  
For the same set of initial conditions and forcing sequence, 
their spectra extracted from runs performed 
using the moving-mesh code Arepo
exhibit an inertial range which extends over more than a decade in $k$.
Similar results were later obtained by \citet[][Figure 27]{ho15}, 
by using for the same test problem a completely different code (Gizmo).
\begin{figure}[t]
\includegraphics[height=9cm,width=7.5cm]{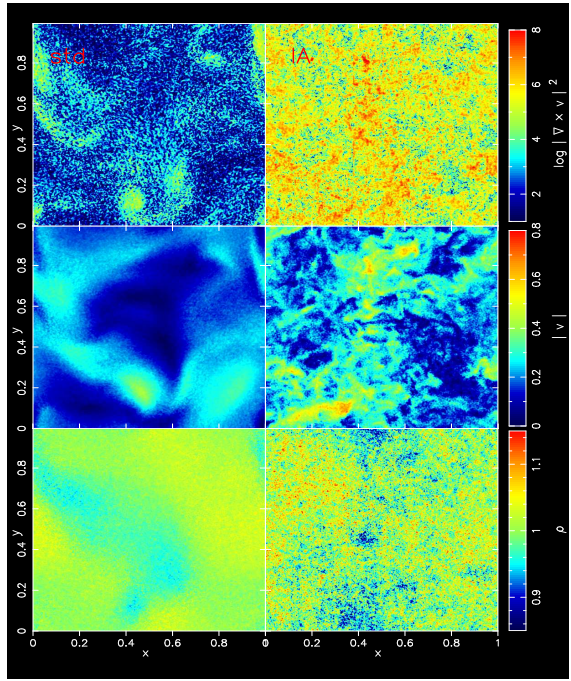}
\caption{ 2D maps of the density (bottom), velocity (middle) and enstrophy(top) 
fields extracted from simulations of driven subsonic turbulence with 
resolution $N^3=256^3$ and at the time $t=25$. The left column is for 
standard SPH and the right column refers to the IA runs.
 The fields are evaluated 
on a grid  of $512^2$ points located at $z=L/2$;  an SPH 
interpolation procedure is used to compute 
field values  from particle quantities at grid points
\label{fig:mapsubs}}
\end{figure}

A similar behavior is found here for the spectra of the IA-CD runs 
depicted in Figure \ref{fig:pwsubs}, which show a dramatic improvement
over the corresponding standard SPH runs.
The spectra exhibit now a much larger inertial range, which 
increases with resolution and for the $N=256$ simulation 
it extends down to $k\sim 200$,
 close to the minimum scale $k_{max}\sim 2 \pi N/5$
estimated by \citet{ho15}.
The spectra are similar to the corresponding ones  shown in 
 Figure 27 of \citet{ho15}, but with fluctuations which stay 
within a factor $\sim$ two for $k \simlt k_{max}$.

{ The velocity power spectra of the IA-AV runs exhibit significant 
differences with respect those of the corresponding IA-CD simulations.
For the sake of clarity we show in Figure \ref{fig:pwsubs} only the 
spectrum of the $N=128$ run. This spectrum is characterized by 
a significant amount of noise, with a departure from its parent IA-CD 
run which begins already at scales above the Nyquist frequency.

This sensitivity of the IA spectra on the adopted AV scheme is 
at variance with what seen in the SPH (TSPH and PSPH) runs of 
\citet{ho15}, in which the impact of AV on spectral behavior 
is not so significant.
We interpret this strong dependence of velocity power spectra on the AV 
 scheme as being due to the effectiveness of the 
IA method in removing gradient errors. This  in turn implies that AV,
which was previously  subdominant \citep{ba12}, is now the main source of 
noise.  The level of noise seen in the IA-AV spectra is then absent in the
spectra of the IA-CD runs,  because of the 
limited amount of AV which is generated by the employed AV switch.
}

Finally, in Figure \ref{fig:mapsubs}  we show 2D maps of the density $\rho$, 
velocity 
$\vec v$ and $|\vec \nabla \times \vec  v |^2$ extracted at $t=50$ from 
simulations with resolution $N^3=128^3$ for both IA and standard SPH runs.
A visual comparison between the maps of the two runs clearly indicates the presence
in the IA simulation of well resolved small-scale features which are absent 
in the corresponding standard SPH map.
These features of the IA maps appear qualitatively very similar to those  
obtained, in their tests on subsonic turbulence, by 
\citet[][Figure 4]{ba12}  
 and by \citet[][Figure 26]{ho15}.

These findings confirm that in SPH simulations, kernel gradient errors 
 play a key role in the modeling of subsonic turbulence, 
and demonstrate how the IA scheme can be profitably used to 
overcome these difficulties, with results which compare well with those
 obtained with other numerical schemes recently proposed.

\subsection{Keplerian disc} \label{subsec:ring}

The cold Keplerian disc problem has been investigated by  
many authors \citep{im02,ca09,cul10,hu14,ho15,sa15,pa16,ho16}.
The test consists of a gaseous disc orbiting around a point-like mass.
The disc has negligible pressure and its self-gravity is 
neglected; the disc is then in equilibrium with the centrifugal 
forces being balanced by the gravity of the central mass.
Because of these conditions, the system is in a steady state and the initial
disc configuration should remain stable as a function of time.
 
For SPH codes this problem is very challenging, since even a small
amount of AV causes a transport of angular momentum leading to 
particle disorder and disc break-up. The problem is particularly 
severe in the inner part of the disc, where the differential rotation 
causes strong shear flows.
\begin{figure*}[!t]
\includegraphics[height=5.8cm,width=16.5cm]{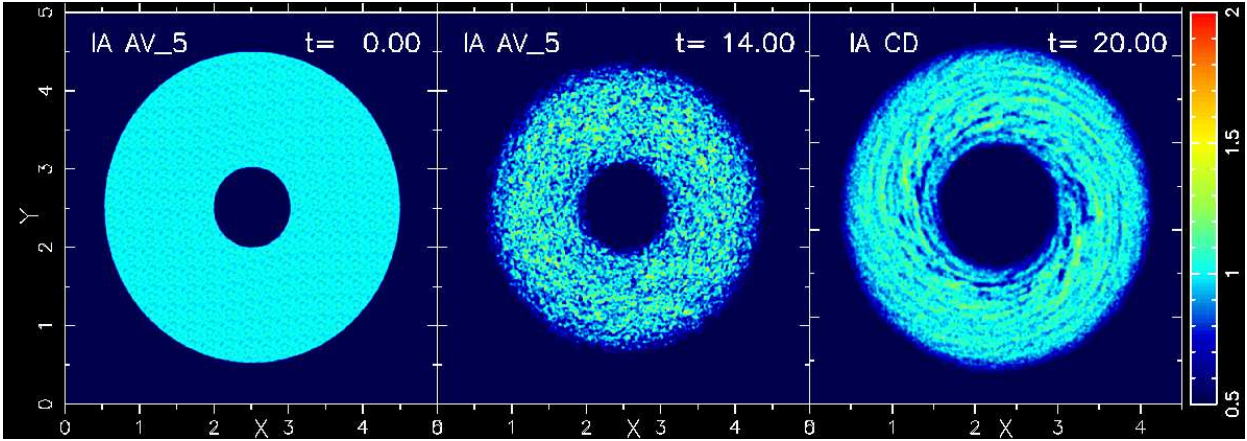}
\caption{ For some of the SPH runs we here show
 at various times (in units of $T=2\pi$), 2D density maps  
of the Keplerian disc.
All of the SPH simulations incorporate the IA scheme,
 the code velocity divergence and vorticity being calculated 
in accordance with the scheme.
Left: initial configuration for a SPH simulation which uses the AV 
setting $\{\alpha_{min},\alpha_{max},l_d\} = \{0.01,1.5,1.0\}$
(see Section \ref{subsec:visco}). Middle : the same simulation but at $t=14$. 
Right : here we show at $t=20$ the disc density for an SPH run in which
the AV scheme being used is that of \citet{cul10}. 
To better discriminate disc structure in this run, the size of the 
computational domain has been reduced to $L_x=L_y=4.5$.
\label{fig:mapdisk}}
\end{figure*}

In SPH, suppression of AV in the presence of shear flows is 
regulated by the Balsara switch (\ref{fdamp.eq}).
Because of this, higher order  velocity gradient estimators must be 
adopted \citep{cul10} in order to prevent to prevent or delay 
 the disc instability in SPH runs. Additionally, because  pressure 
forces are very small,  
 zeroth-order errors in hydrodynamic forces can also 
play a role in developing instabilities.
It is therefore interesting to investigate the performance of the 
proposed IA scheme with the cold Keplerian disc problem.

We implement our initial conditions in a manner similar to that of 
previous authors \citep{ho15,sa15}.
More specifically, we set-up the disc in a three dimensional 
periodic domain with boundaries defined by 
$\{-2.5,-2.5,0\} \leq x,y,z < \{2.5,2.5,0.25\}\equiv\{L_x/2,L_y/2,L_z\}$. 
Within this domain the disc density satisfies 

\be
\rho(r)= \left\{
\begin{array}{ll}
1   &   0.5\leq r\leq 2 \\
0  &   2 < r~,  \\
\end{array}
\right.
\label{rhokd.eq}
\ee

where $r$ is the cylindrical radius $r=\sqrt{ x^2+y^2}$.  

Unlike in previous settings \citep{ho15,sa15} here the disc edges are 
unsmoothed,  moreover we put 
in the $x-y$ plane a small empty zone around the disc  to avoid 
border effects so as to mimic vacuum boundary conditions.
The gas initially has a very small  constant pressure, $P=10^{-6}$,
and index $\gamma=5/3$.

For the central point mass we set  $GM=1$ and the gas is 
subject to a static gravitational acceleration $\vec a =- \vec \nabla \Phi$ ,
where $\Phi=-(r^2+\epsilon^2)^{-1/2} $ is the softened potential.
We introduce a softening parameter, $\epsilon=0.25$, to avoid 
diverging accelerations for those particles that during  the 
simulations escape the initial disc configuration and approach the origin 
at $r=0$. The initial particle rotational velocity is then 
$V_{\phi}= r  (r^2+\epsilon^2)^{-3/4}$ and for the Keplerian orbital period 
$T=2\pi r^{3/2}$ we choose as reference value that at $r=1$.
Hereafter we express time  in units of $T= 2\pi$.

We implement the initial condition set-up according to the 
following procedure. We first construct a uniform glass-like distribution
of $256^2 \times 16$ particles  inside a  parallelepiped of side lengths
 $4 \times 4\times 0.25$.  
This is done by creating  $16\times 16 $ replicas of a root unit cube of 
$16^3$ glass-like particles along each $x$ and $y$ axis. 
and then rescaling the parallelepiped.
Finally, we only keep those particles whose $(x,y)$ coordinates 
satisfy the conditions given in Equation (\ref{rhokd.eq}).
 The final number of particles used in the simulations is then 
$N=16 \cdot 256^2 \pi/4\sim 8.2 \cdot 10^5$. The velocities of the 
particles are initialized consistently with  their position.
In all of the SPH runs we use the same initial conditions set-up and the kernel
($W4$) with $N_n=200$ neighbors. The simulations are evolved up to a 
maximum time $t=20$.

It must be stressed that the capability of the code to follow the disc
orbits depends sensitively of how the initial particle configuration
has been chosen in order to minimize the growth of 
numerical instabilities \citep{ca09}. 
In this aspect, after several tests, it has been found that the most 
stable discs are obtained when a glass-like particle distribution  is used to 
realize the density set-up (\ref{rhokd.eq}).
 
For some of the simulations performed, we show in Figure \ref{fig:mapdisk} 
density maps  of the simulated discs at various times. The maps are 2D slices 
calculated on an $800\times 800$ grid located at $z=L_z/2$ in the 
simulation domain.

We do not show here results from disc simulations performed with  the 
standard SPH implementation; these are in line with previous findings and  
the disc is found to be subject to disruption after few orbits ($t\sim 2-3$).
Introducing the IA scheme significantly improves the code 
capability for evolving the disc, which now can be followed  up to 
$t \sim 15 $ before it begins to degrade ( Figure \ref{fig:mapdisk}, 
left and middle panels).

This SPH simulation (IA-AV$_5$) was performed by setting 
$\{\alpha_{min},\alpha_{max},l_d\} = \{0.01,1.5,1.0\}$ 
 for the AV parameters, this choice being indicated 
with the notation  AV$_5$ in  previous calibration tests \citep{va11}.
With respect to the set of AV parameters adopted in the other tests 
performed here (AV$_2$), the setting AV$_5$ is characterized by a very 
low floor value ($\alpha_{min} = 0.01$)
 and  the shortest possible decay timescale ($l_d = 1.0$).
This choice of AV parameters in SPH runs, improves the disc stability though 
not in a significant way, with disc break-up occurring at $t\sim12$ when 
the setting AV$_2$ is used. 

Moreover, it must be stressed that in the IA runs the velocity 
divergence and vorticity are calculated from a velocity gradient matrix,
in accord with the adopted IA scheme. The results presented here are however 
still valid if a 
standard SPH estimator is used in the calculation of the velocity gradients, 
with disc stability being affected only marginally.
This shows that errors in hydrodynamic forces  are dominant 
in determining  disc stability,  with respect to low-order errors  affecting 
 the shear viscosity limiter (\ref{fdamp.eq}) 

A significant improvement in disc stability  is obtained by replacing 
the time-dependent AV scheme of Section \ref{subsec:visco} with the 
improved method proposed by \citet{cul10}, which is still based on the 
\citet{mm97} scheme but has a better shock indicator and 
a more accurate AV limiter.

An IA-SPH simulation (IA-CD) performed by incorporating the  new AV scheme  
shows that 
the small amount of AV, which still affected disc evolution in the 
IA-AV$_5$ run, is now removed and the disc structure is now stable up to 
20 orbits.
The disc density map of the IA-CD run is shown at $t=20$ in the right panel of 
Figure \ref{fig:mapdisk}, 
and can be compared with the 
corresponding  density maps shown in Figure 6 of \citet{ho15}. 

A comparison between the two suites of simulations is possible because 
we adopt here the same initial condition set-up. In previous 
papers \citep{hu14,be16} the Gaussian ring version of the problem
has been used to test  new versions of SPH, but we expect our conclusions
to remain unaffected by  our choice of initial conditions.

To summarize, the results presented here for the Keplerian disc problem 
demonstrate that for an SPH code  both errors in gradient accuracy and the 
level of AV contribute to disc stability, with the former having a much greater
impact.


\section{Conclusions} \label{sec:conc}


In this paper we have investigated the performance of an 
improved version of the standard SPH formulation, 
in which an integral approach  is used to strongly reduce 
zeroth-order errors in gradient estimates.

The IA method has been proposed and tested in a variety of hydrodynamical 
test problems \citep{ga12,ro15}, but its most significant applications
are in the simulation of subsonic flows. 
In the low Mach number regime the difficulties 
of standard SPH have been found to be particularly severe 
\citep{ba12,de12,ga12,na12,va12,ho15} and gradient accuracy
is a key prerequisite for accurate modeling the fluid dynamics.

Given the advantages of a numerical hydrodynamical scheme based on a 
Lagrangian formulation (for instance its natural
resolution adaptativity) it is therefore crucial to assess the capability 
of the proposed IA-SPH scheme to handle subsonic flows.
Moreover, the IA method retains the fully
conservative nature of the Lagrangian SPH scheme, unlike previous
attempts aimed at removing zeroth-order gradient errors present in SPH.

To evaluate code performance we have analyzed results  
from a suite of simulations of hydrodynamical test problems, 
performed using both the IA and standard SPH formulations.
We also contrast the accuracy of the results with that produced 
by new numerical schemes \citep{sp10,ba12,ho15}, against which standard SPH 
has been found clearly inadequate.
Our main conclusions are  as follows.

For the Gresho-Chan vortex problem it is well known \citep{read12,de12,hu14,ho15} that 
 standard SPH is heavily affected by the $E_0$ error and the code performances 
are very poor. On the contrary, the IA formulation leads to much better 
behavior, with the results of Section \ref{subsec:gre}
being in line with those obtained  by 
other numerical schemes \citep{sp10,ho15}.

The resolution study displayed in  Figure \ref{fig:vtxL1}  shows 
for the $L1$ velocity error a gain in accuracy by a factor $\sim10$ over 
standard SPH. Moreover, the validity of the approach is confirmed even 
in the regime of very cold flows. This is demonstrated by the velocity 
profiles of Figure \ref{fig:vtxlow}, in which the code is shown to be able to 
reproduce the analytic solution for the azimuthal velocity  
down to $M=0.02$.

The results of Section \ref{subsec:kh} on KH tests also indicate how 
zeroth-order errors present in SPH affect the growth of KH instabilities
and the effectiveness of the IA scheme in reducing these errors. Nonetheless, 
in the very low subsonic regimes, the IA method shows, at a fixed resolution, 
a progressively reduced capability to follow the development of the instability. 
This is not surprising since, for the chosen settings, by reducing the 
Mach number the perturbation amplitude is also reduced and it becomes 
progressively more challenging, keeping the resolution fixed, 
 to simulate the KH instability when $M\simlt0.1$.

In this respect, the heuristic arguments used in Section \ref{subsec:kh}  
to derive the necessary resolution give a lower limit for $N$ which in 
any case should be taken with caution, with the required value probable 
being much higher.
In fact, to the author's knowledge, KH simulations with very low Mach numbers 
($M\simlt0.1$) have not previously been undertaken in the literature 
and it would be interesting to 
compare the findings of Section  \ref{subsec:kh} with the behavior
of a mesh-based code in these regimes.
   
The good performances of the IA formulation are confirmed by the results 
of Section \ref{subsec:subs}
 on simulations of driven  subsonic  turbulence, for which the failure 
 of standard SPH to properly model this problem has been debated by 
various authors \citep{ba12,pr12b,ho13,ho15,zu15}.
Simulations performed  employing the new scheme produce velocity spectra 
in better agreement with the Kolgomorov law and exhibit an inertial range 
which now covers nearly a decade (Figure \ref {fig:pwsubs}).
Here again we see how the results, for the same initial setting and resolution, 
 do not differ significantly from those 
produced by other codes \citep{ba12,ho15}.

In the Keplerian disc problem, suppression of numerical viscosity is a critical 
factor for achieving stable evolution. However the results of Section
\ref{subsec:ring} also show  errors in hydrodynamical forces having
a significant impact on disc stability. This suggests that the instabilities 
leading to disc disruption are sourced by a combination of these two factors.
To successfully simulate this test problem, an SPH code must then be 
necessarily based on both the IA scheme and the improved AV switch of
 \citet{cul10}.

The use of an IA scheme within an SPH framework also raises the issue of revisiting
 the choice of the kernel in SPH simulations. 
As discussed  in Section \ref{sec:ker}, the introduction of Wendland kernels stems
 from the necessity of avoiding pairing instability. 
But this problem arose from the need to reduce zeroth-order errors present in 
standard SPH, which are absent or very small in the IA scheme. Therefore, if one
adopts the IA-SPH framework, one can resort to the use of the 
$M_5$ or $M_6$ splines in place of the Wendland kernels. This choice  is
motivated, for the same  number of neighbors, by the  better accuracy of the 
B-splines in estimating densities, when contrasted against the Wendland kernels
 (see Section \ref{sec:ker}).

To summarize, the results of our tests demonstrate that by incorporating 
the IA method in standard SPH, the zeroth-order errors in the momentum 
equations are drastically reduced, with significant improvements in the 
performance of the new code. 

These results are particularly significative given the importance of
 subsonic flows  in many astrophysical 
problems. For example, in galaxy clusters subsonic turbulence  adds a 
contribution to the 
intracluster medium pressure, thus biasing cluster mass estimates and in turn 
affecting the use of clusters as cosmological probes 
\citep[][and references cited therein]{br15}.

We thus conclude that the new IA-SPH scheme, being based on a  Lagrangian 
formulation, can be profitably used
 in those simulations of subsonic astrophysical 
flows in which the shortcomings of standard SPH prevented full exploitation
of its resolution adaptativity and conservation properties.

\acknowledgments

{The author would like to thank 
D. Price 
for kindly  supplying the driving routine for subsonic turbulence,  
on which the code used in Section \ref{subsec:subs} is based.
The authors also thanks the anonymous referee for constructive comments that
substantially improved this paper.
This work has been supported by the MIUR PRIN 2010/2011
``The Dark Universe and the Cosmic Evolution of Baryons: 
from Current Surveys to Euclid''.
The computations of this paper were performed using the Ulisse cluster at SISSA
and the Galileo cluster at CINECA (Italy), under a SISSA-CINECA agreement}

\clearpage



\clearpage



\begin{thebibliography}{}

\bibitem[Abell (2011)]{ab11}
Abel, T., \ 2011, \mnras, 413, 271

\bibitem[Agertz et al. (2007)]{ag07}
{Agertz}, O., {Moore}, B., {Stadel}, J., {Potter}, D.,
{Miniati}, F., {Read}, J., {Mayer}, L., {Gawryszczak}, A., 
{Kravtsov}, A., {Nordlund}, {\AA}, {Pearce}, F., {Quilis}, V.,  
{Rudd}, D., {Springel}, V., {Stone}, J., {Tasker}, E.,  
{Teyssier}, R., {Wadsley}, J. \& {Walder}, R., 2007, \mnras , 380, 963


\bibitem[Aguilar et al. (2011)]{ag11}
Aguilar, J. C., Berriel-Valdos, L. R. \& Aguilar, J. F., 2011, 
 in 22nd Congress of the International Commission for Optics, 
Proceedings of SPIE, vol. 8011, 801186


\bibitem[Balsara (1995)]{ba95}
Balsara, D.,  1995, J. Comp. Phys., 121, 357

\bibitem[Bartosch (2001)]{ba01}
Bartosch, L.,  2001, Int. Jour. of Modern Physics, 12, 851

\bibitem[Bauer \& {Springel} (2012)]{ba12}
Bauer, A. \& {Springel}, V.,  2012,  \mnras , 423, 2558


\bibitem[Beck et al. (2016)]{be16}
 {{Beck}, A. M., {Murante}, G.,  {Arth}, A.,  {Remus}, R.-S.,  
{Teklu}, A. F.,  {Donnert}, J. M. F.,  {Planelles}, S.,
  {Beck}, M. C., {Foerster}, P.,  {Imgrund}, M.,  {Dolag}, K.  \& 
 {Borgani}, S.}, 2016, \mnras, 455, 2110

\bibitem[Biffi \& Valdarnini (2015)]{bi15}
Biffi, V. \& Valdarnini, R., 2015, \mnras, 446, 2802

\bibitem[B{\o}rve et al. (2004)] {borv04}
B{\o}rve, S., {Omang}, M.  \& {Trulsen}, J., \ 2004, \apjs,  153, 447


\bibitem[Brookshaw (1985)]{br85}
Brookshaw, L., \ 1985, Proc. of the Astrn. Society of Australia, 6, 207

\bibitem[Br{\"u}ggen \& {Vazza} (2015)]{br15}
Br{\"u}ggen, M. \& {Vazza}, F., \ 2015, 
Astrophysics and Space Science Library, 407, 599, 
A. Lazarian et al. (eds.),  Springer-Verlag




\bibitem[Bryan et al. (2014)]{br14}
Bryan, G.~L. ,{Norman}, M.~L., {O'Shea}, B.~W., {Abel}, T.,  
  {Wise}, J.~H., {Turk}, M.~J., Reynolds, D.~R., {Collins}, D.~C.,  
 2014, \apjs, 211, 19

\bibitem[Cartwright et al. (2009)]{ca09}
{Cartwright}, A., Stamatellos, D. \& {Whitworth}, A.~P., \ 2009, \mnras, 
395, 2373 

\bibitem[{Cha} et al.  (2010)]{ch10}
{Cha}, S.-H., {Inutsuka}, S.-I. \& {Nayakshin}, S., \ 2010, \mnras, 403, 1165


\bibitem[Cullen \& Dehnen (2010)]{cul10}
Cullen, L. \& Dehnen, W., 2010, \mnras, 408, 669

\bibitem[Dehnen \& Aly (2012) ]{de12}
Dehnen, W. \& Aly, H.,  2012,   \mnras, 425, 1068

\bibitem[Dilts (1999)]{di99}
Dilts, G., 1999, Int. J. Numer. Methods Eng., 44, 1115

\bibitem[Duffell \& {MacFadyen} (2011)]{du11}
Duffell, P.~C. \& {MacFadyen}, A.~I., 2011, \apjs, 197, 15

\bibitem[Eswaran \& Pope  (1988)]{es88}
Eswaran, V. \& Pope, S.~B., 1988, Computers \& Fluids, 16, 257


\bibitem[Fryxell et al. (2000)]{fr00}
Fryxell, B., Olson, K., Ricker, P., Timmes, F.~X.,
        Zingale, M., Lamb, D.~Q., MacNeice, P., Rosner, R.,
        Truran, J.~W. \& Tufo, H., \ 2000 \apjs, 131, 273


\bibitem[Garc{\'{\i}}a-Senz et al.  (2012)]{ga12}
{Garc{\'{\i}}a-Senz}, D., Cabez{\'o}n, R.~M. \& {Escart{\'{\i}}n}, J.~A.,
  2012, \aap, 538, A9

\bibitem[{Garc{\'{\i}}a-Senz et al. }  (2014)]{ga14}
{Garc{\'{\i}}a-Senz}, D., Cabez{\'o}n, R.~M., {Escart{\'{\i}}n}, J.~A. \& 
Ebinger, K.,   2014, \aap, 570, A14

\bibitem[Gingold  \& Monaghan  (1977)]{gm77}
Gingold,  R.~A. \& Monaghan,  J.~J.,  \ 1977, \mnras, 181, 375


\bibitem[Gresho \& Chan (1990)]{gr90}
Gresho, P.~M. \& Chan, S.~T., 1990, IJNMF, 11, 621

\bibitem[Hernquist \& Katz (1989)]{hk89}
Hernquist, L. \& Katz, N.,  1989 \apjs, 70, 419


\bibitem[He{\ss} \& {Springel} (2010)]{hs10}
He{\ss}, S. \& {Springel}, V., \ 2010 , \mnras, 406, 2289


\bibitem[Hopkins (2013)]{ho13}
Hopkins, P. F., 2013, \mnras , 428, 2840

\bibitem[Hopkins (2015)]{ho15}
Hopkins, P. F., 2015, \mnras , 450, 53

\bibitem[Hosono et al. (2016)]{ho16}
Hosono, N., Saitoh, T.~R. \& Makino, J., 2016, \apjs, 224, 32

\bibitem[Hu et al. (2014)]{hu14}
{{Hu}, C.-Y.,  {Naab}, T., {Walch}, S., {Moster}, B. P. \&
     {Oser}, L.}, 2014, \mnras, 443, 1173

\bibitem[Imaeda \& Inutsuka (2002)]{im02}
{Imaeda}, Y. \&  {Inutsuka}, S.-i., \ 2002, \apj,  569, 501

\bibitem[ Inutsuka (2002)]{I02}
Inutsuka, S.-I., \ 2002, J. Comp. Physics,  179, 238



\bibitem[Junk et al. (2010)]{ju10}
{Junk}, V., {Walch}, S., {Heitsch}, F., Burkert, A.,
        {Wetzstein}, M., {Schartmann}, M. \& {Price}, D.,
 2010, \mnras, 407, 1933

\bibitem[Kawata et al. (2014)]{ka14}
{{Kawata}, D., {Okamoto}, T.,  {Gibson}, B.~K., {Barnes}, D.~J. \& 
	{Cen}, R.}, 2013, \mnras, 428, 1968

\bibitem[Kravtsov et al. (1997)]{kr97}
Kravtsov, A.~V., Klypin, A.~A. \& Khokhlov, A.~M., \ 1997, \apjs, 111, 73


\bibitem[Liu et al. (2003)]{li03}
Liu, M. B., Liu, G. R. \& Lam, K. Y., 2003, J. Comput. Appl. Math., 155, 263

\bibitem[Lucy (1977)]{lu77}
Lucy, L.~B., \ 1977 , Astr. Journal, 82, 1013


\bibitem[{McNally} et al. (2012)]{na12}
{{McNally}, C. P., {Lyra}, W. \& {Passy}, J.-C.},  2012,  \apjs, 201, 18

\bibitem[Miczek et al. (2015)]{mic15}
{{Miczek}, F. and {R{\"o}pke}, F.~K. and {Edelmann}, P.~V.~F.}, 2015, \aa, 
576, A50



\bibitem[Mitchell et al. (2009)]{mi09}
{Mitchell}, N.~L., {McCarthy}, I.~G., {Bower}, R.~G.,
     {Theuns}, T. \& {Crain}, R.~A., 2009 , \mnras, 395, 180


\bibitem[Monaghan (1997)]{mo97}
Monaghan, J. J.,  1997,  J. Comp. Physics, 136, 298

\bibitem[Monaghan (2005)]{mo05}
Monaghan, J.~J.,  2005, Reports on Progress in Physics, 68, 1703


\bibitem[Morris (1996)]{mo96}
Morris, J. P., 1996, PASA, 13 ,97

\bibitem[Morris \& Monaghan  (1997)]{mm97}
Morris, J. P. \& Monaghan, J. J.,  1997, J. Comp. Physics, 136, 41


\bibitem[Murante et al. (2011)]{mu11}
 {{Murante}, G., {Borgani}, S., {Brunino}, R. \& {Cha}, S.-H.},
 \ 2011, \mnras, 417, 136



\bibitem[Norman \& Bryan (1999)]{no99}
Norman, M.~L. \& Bryan, G.~L., 1999,
in Numerical Astrophysics, Astrophysics and Space Science Library Vol. 240,
ed. S.~M.~Miyama, K.~Tomisaka, \& T.~Hanawa (Kluwer, Boston) , 19

\bibitem[{Norman} (2005)]{no05}
{Norman}, M.~L., \ 2005, {{The Impact of AMR in Numerical Astrophysics and
 Cosmology},} in {Adaptive Mesh Refinement -- Theory and Applications.
 Springer, Berlin, New York}, vol.~41 of {Plewa T., Linde T., Weirs
 V.G, eds, Lecture Notes in Computational Science and Engineering}, p. 413.


\bibitem[{O'Shea} et~al. (2005)]{os05}
{O'Shea}, B.~W., {Bryan}, G., {Bordner}, J., {Norman}, M.~L., {Abel}, T.,
  {Harkness}, R. \& {Kritsuk}, A., 2005,
 {{Introducing Enzo, an AMR Cosmology
  Application},} in {Adaptive Mesh Refinement -- Theory and Applications,
  ed.~T.~Plewa, T.~Linde, V.G.~Weirs (Berlin; New York: Springer)},
  vol.~41 of {Lecture Notes in Computational Science and Engineering}, p.
  341.


\bibitem[Pakmor et al. (2016)]{pa16}
 {{Pakmor}, R., {Springel}, V., {Bauer}, A., {Mocz}, P.,  
 {Munoz}, D. J., {Ohlmann}, S. T., {Schaal}, K. \& {Zhu}, C. }, 
 2016,  \mnras, 455, 1134

\bibitem[Price (2008)]{pr08}
Price, D.J.,  2008, J. Comp. Phys. , 227, 10040


\bibitem[Price \& Federrath (2010)]{pr10}
Price, D. J. \& Federrath, C.,  2010, \mnras, 406, 1659

\bibitem[Price (2012)]{pr12}
Price, D. J.,  2012, J. Comp. Phys., 231, 759

\bibitem[Price (2012b)]{pr12b}
Price,  D. J.,  2012b,  \mnras, 402, L33

\bibitem[{Read} et al. (2010)]{read10}
{{Read}, J.~I., {Hayfield}, T. \& {Agertz}, O.}, 2010, \mnras, 405, 1513

\bibitem[{Read} \& {Hayfield} (2012)]{read12}
{Read}, J.~I. \& {Hayfield}, T.,  2012,  \mnras, 422, 3037


\bibitem[Ritchie  \& Thomas  (2001)]{ri01} 
Ritchie, B. W. \& Thomas, P. A., 2001, \mnras, 323, 743

\bibitem[{Robertson} et al. (2010)]{rb10}
 {{Robertson}, B.~E., {Kravtsov}, A.~V., {Gnedin}, N.~Y.,
        {Abel}, T. \& {Rudd}, D.~H.}, \ 2010, \mnras, 401, 2463


\bibitem[Rosswog (2009)]{ro09}
Rosswog, S., 2009, New  Astron. Reviews, 53, 78

\bibitem[Rosswog (2015)]{ro15}
Rosswog, S., 2015, \mnras , 448, 3628

\bibitem[Saitoh \& Makino (2013)]{sa13}
Saitoh, T.~R. \& Makino, J., \ 2013, \apj, 768, 44


\bibitem[Schaal et al. (2015)]{sa15}
{Schaal}, K., {Bauer}, A., {Chandrashekar}, P., {Pakmor}, R., 
 {Klingenberg}, C. \& {Springel}, V.,   2015, \mnras, 453, 4278


\bibitem[Springel \& Hernquist (2002)]{SH02}
Springel, V. \& Hernquist, L.,  2002, \mnras, 333, 649

\bibitem[Springel (2005)]{sp05}
Springel, V., \ 2005, \mnras, 364, 1105


\bibitem[Springel (2010)]{sp10}
Springel, V.,  2010, \mnras, 401, 791


\bibitem[Stone \& Norman (1992)]{st92}
Stone, J.~M. \& Norman, M.~L., \ 1992, \apjs , 80, 753

\bibitem[Stone et al. (2008)]{sto08}
{Stone, J.~M., Gardiner}, T.~A., {Teuben}, P., {Hawley}, J.~F.  \&
   {Simon, J.~B.},  2008, \apjs, 178, 137

\bibitem[Tasker et al. (2008)]{ta08}
{Tasker}, E.~J., {Brunino}, R., {Mitchell}, N.~L.,
    {Michielsen}, D., {Hopton}, S., {Pearce}, F.~R., {Bryan}, G.~L. \&
     {Theuns}, T., 2008, \mnras, 390, 1267



\bibitem[Teyssier (2002)]{te02}
Teyssier, R., \ 2002, \aap, 385, 337

\bibitem[Valcke et al. (2010)]{vrd10}
{{Valcke}, S., {de Rijcke}, S., {R{\"o}diger}, E. \& {Dejonghe}, H.},
 \ 2010, \mnras,  408, 71

\bibitem[Valdarnini (2011)]{va11}
Valdarnini, R., 2011, \aap, 526, A158

\bibitem[Valdarnini (2012)]{va12}
Valdarnini, R., 2012, \aap, 546, A45

\bibitem[Wadsley et al.   (2004)]{wa04}
{Wadsley}, J.~W., {Stadel}, J. \& {Quinn}, T., \ 2004 , New Astr., 9, 137

\bibitem[Wadsley et al.  (2008)]{wa08}
{Wadsley}, J.~W., {Veeravalli}, G. \& {Couchman}, H.~M.~P., 2008 ,
\mnras , 387, 427


\bibitem[Wendland (1995)]{we95}
Wendland, H., 1995, Adv. Comp. Math., 4, 389


\bibitem[Zhu et al. (2015)] {zu15}
 {{Zhu}, Q., {Hernquist}, L. \& {Li}, Y.}, 2015, \apj , 800, 6


\end{thebibliography}
\end{document}